\documentclass[10pt,twocolumn,pra,showpacs,amsmath,
amssymb,aps,floatfix,superscriptaddress,longbibliography]{revtex4-1}

\usepackage{bbm}
\usepackage{empheq}
\usepackage[utf8]{inputenc}
\usepackage{enumerate}

\usepackage{color}

\usepackage{amsthm}
\usepackage{mathrsfs}
\usepackage{textcomp}

\usepackage{graphicx}
\usepackage{epstopdf}

\usepackage[breaklinks=true]{hyperref}
\usepackage{breakcites}

\usepackage{todonotes}

\newcommand{\id}{\mathbbm{1}}                %Identity
\newcommand{\tr}[1]{\operatorname{Tr}\left[ {#1} \right]} %Trace
\newcommand{\ket}[1]{\left|#1 \right \rangle\! \vphantom{\left( #1 \right)^A}} %ket
\newcommand{\bra}[1]{\left\langle #1 \right | \vphantom{\left(#1\right)^A}} %bra
\newcommand{\sys}{\mathcal{S}}  %\sss --> Snapshot SubSpace
\newcommand{\ev}{\text{event}}  %\sss --> Snapshot SubSpace
\newcommand{\env}{\mathcal{E}} %\hk --> K SubSpace

\newcommand{\rank}[1]{\relax\ifmmode\operatorname{rank}#1\else rank-$#1$\fi}

\begin{document}

\title{A single-world consistent interpretation of quantum mechanics from fundamental time and length uncertainties}

\author{Rodolfo Gambini}
\affiliation{Instituto de F\'{\i}sica, Facultad de Ciencias,
  Universidad de la Rep\'ublica, Montevideo, Uruguay}
\author{Luis Pedro Garc\'ia-Pintos}
\affiliation{Department of Physics, University of Massachusetts, Boston, MA
02125, USA}
\author{Jorge Pullin}
\affiliation{Department of Physics and Astronomy, Louisiana State
  University, Baton Rouge, LA 70803, USA}

\date{\today}

\begin{abstract}
Within ordinary ---unitary--- quantum mechanics there exist global
protocols that allow to verify that no definite event ---an outcome to
which a probability can be associated--- occurs. Instead, states that start in a coherent superposition over possible outcomes always remain as a superposition.
We show that, when taking into account fundamental errors in measuring
length and time intervals, that have been put forward as a consequence
of a conjunction of quantum mechanical and general relativity arguments, 
there are instances in which 
such global protocols no longer allow to distinguish whether the state is in a superposition or not. 
All predictions become identical as if one of the outcomes occurs, with probability determined by the state.
We use this as a criteria to define events, as put forward in the
Montevideo Interpretation of Quantum Mechanics. We analyze in detail the occurrence of events in the paradigmatic case of a particle in a superposition of two different locations.
%Thereafter, we revisit a theorem by Frauchiger and Renner showing that having a consistent single world description of the universe is incompatible with quantum theory, and we show that our approach provides the most economical way out of the limitations imposed by it.
%The Montevideo Interpretation thus provides a consistent single world picture of the universe. 
We argue that 
our approach 
%the Montevideo Interpretation 
provides a consistent (C) single-world (S) picture of the universe, thus
allowing an economical way out of the limitations imposed by a recent theorem by Frauchiger and Renner showing that having a self-consistent
%\textcolor{blue}{theory where agents observe single outcomes}
single-world description of the universe
is incompatible with quantum theory.
In fact, the main observation of this paper may be stated as follows: 
If quantum mechanics is extended to include gravitational effects to a QG theory, then QG, S, and C are satisfied. 
\end{abstract}

\maketitle

% \vspace{10pt}
% \noindent \emph{Introduction and setting}
% \vspace{10pt}

%
%XXXXX for the abstract:
%
%XXXXX
%Environmental decoherence has been put forward as a means to explain the productions of quantum events, providing a criteria that, for all practical purposes, looks like a collapse of the wavefunction. 
%However, this route does not lead to definite events, one of the reasons being that there exist global protocols that, under unitary evolution allow to distinguish the occurrence of events in quantum theory. 
%
%XXXXX?
%
%XXXXX
A crucial feature of quantum mechanics is that, in general, the state of a quantum system cannot be interpreted as a classical probability distribution. 
In particular, when a closed system $\sys$ is found in a coherent superposition of a set of orthonormal states $\{ \ket{\varphi_j }\}$,
\begin{equation}
\label{eq:superpositionstate}
\rho_\sys = \sum_{jk} \alpha_j \alpha^*_k \ket{\varphi_j}\!\bra{\varphi_k}, 
\end{equation}
with $\sum_j |\alpha_j|^2 = 1$, one cannot view $\sys$ as being described by a classical probability distribution over the states $\{ \ket{\varphi_j }\}$.
This is in stark contrast to the case of a system in a state
\begin{equation}
\label{eq:classicalstate}
\rho_\sys^{\ev} = \sum_{j} |\alpha_j|^2 \ket{\varphi_j}\!\bra{\varphi_j}.
\end{equation}
%which can be interpreted as a classical probability distribution ; 
For the latter, all subsequent dynamics and predictions are exactly as if the system had been described by a classical probability distribution over the states $\{ \ket{\varphi_j }\}$.
%The difference between a system in state $\rho_\sys$ and one in state $\rho_\sys^{\ev}$ is experimentally manifested when measuring the system on a different basis, in which case interference terms become relevant.
%\textcolor{red}{RG: no entiendo, en la base varphi tambien es manifiesta, solo los elementos diagonales coinciden, lo que queres decir es que las probabilidades de medir un autovalor asociado a una base diferente los terminos  diagonales se hacen relevantes. XXX Pensar}
Note that we demand
$\sys$ to be a closed system. 
In fact, condition~\eqref{eq:classicalstate} is not sufficient for open systems, for instance for systems that are entangled with other systems, in which case state~\eqref{eq:classicalstate} only represents a reduced density matrix.

The measurement postulate in quantum mechanics \emph{presumes} that, on performing a projective measurement on the basis $\{ \ket{\varphi_j }\}$, a definite transition from $\rho_\sys$ %in~\eqref{eq:superpositionstate} 
to $\rho_\sys^\ev$ %in~\eqref{eq:classicalstate} 
occurs. 
Thereafter, when a particular outcome is observed, the state of the
system is updated to reflect the observed outcome, much the same
way as is done when one replaces one's knowledge after observing
`heads' on a coin toss. This change from $\rho_\sys$ to
$\rho_\sys^\ev$ through a measurement is in stark contrast to what
occurs in classical physics, where the role of a measurement is merely
that of information acquisition. It is also in stark contrast with a
unitary evolution of the system $\sys$. Hence, our stance is that a complete understanding of quantum theory rests on adequately characterizing what does and does not constitute a measurement.

In this paper we solve the above conundrum by providing a definite criterion for the physical means by which a transition from an arbitrary quantum state $\rho_\sys$ of a closed system to a classically interpretable state $\rho_\sys^\ev$ occurs. We call such transition an \emph{event}.

Note that, within realist interpretations of quantum theory, the
problem of explaining the transition
\emph{no~event}~$\rightarrow$~\emph{event} does not pertain to a
particular view about the ontological status assigned to the
wavefunction.  As long as one believes that situations exist in which
physical predictions for the system change when going from $\rho_\sys$
to $\rho_\sys^\ev$, then a criterion is needed to explain how, and
under which conditions, this change occurs.  Such a physical change
occurs even in models in which the state is assumed to have an
epistemological nature~\cite{spekkens2007,Leifer2014}. Indeed, in this
article we  do not need to take any stance on the status of the wavefunction, it
suffices to say that it is the best tool available to predict future
dynamics.  Moreover, while our derivations are in the Schr\"odinger
picture, the state is not a fundamental entity, and the arguments
could be done purely in terms of observables in the Heisenberg picture
instead.

%As we will argue in the Discussion, the ontology in our
%interpretation is based on the occurrence of events, and the
%properties associated to them~\cite{GGPPaxiomatic2011}.

In previous works we showed how fundamental uncertainties in the measurement of time and spin components, stemming from quantum mechanical and general relativistic arguments, can be used to surmount such problem in a particular model of a spin interacting with a spin bath environment~\cite{GGPPundecidability2010}, and provided a criterion for the production of events~\cite{GGPPcriterionevents2011}. 
This set the foundations for a realist interpretation of quantum theory: the Montevideo Interpretation~\cite{GPfaq2009,GGPPaxiomatic2011,
gambini2015montevideo}. 
The main point of this paper, however, does
not depend on the Montevideo Interpretation but just reinforces one of
its basic observations.

Here we extend the solution to a general class of global protocols that apply to any decoherence model.
In this way, we provide a criterion that works in much more general settings. Our analysis also permits to provide estimates for bounds on the time of event occurrence in the paradigmatic situation of a delocalized particle in a coherent superposition over two different locations.

The structure of the paper is as follows. Section~\ref{sec:environmentaldecoherence} reviews the process of environmental decoherence, and its relevance to measurement processes.
In Section~\ref{sec:globalprotocol} we introduce global protocols that, within unitary quantum mechanics, allow to verify that no definite event can occur.
Section~\ref{sec:fundamentaluncertainties} is devoted to fundamental time and length uncertainties, that have been argued in the past to be a consequence of combining general arguments from quantum theory and from general relativity.
The core of the article can be found in Sections~\ref{sec:events} and~\ref{sec:definiteevents}, where we show the physical process that leads to production of events, and that such production is definite rather than for all practical purposes, respectively.
In Section~\ref{sec:frauchigerrenner} we explain how this results in a consistent single-world description of the universe.
We end in~\ref{sec:discussion} with a discussion.

\section{Evolution of non-isolated systems}
\label{sec:environmentaldecoherence}

It can be argued that a breakthrough on the understanding of measurements in quantum mechanics has been made in the past decades from acknowledging that measurements are performed by devices that interact with an environment~\cite{Zeh1970}.
In this section we review what typically happens in cases of a system interacting with an environment.

Consider that $\mathcal{S}$ interacts with an environment $\mathcal{E}$. For simplicity we assume that both are initially in a pure state, $\rho_\sys(0)$ (given by Eq.~\eqref{eq:classicalstate}) and $\ket{E_r}\!\bra{E_r}$ respectively, and that they are uncorrelated. 
That is, the joint initial state is
\begin{equation}
\rho_{\sys \env} (0) = \rho_\sys (0) \otimes \ket{E_r}\!\bra{E_r}.
\end{equation}
Following the standard setup for the interaction of a measuring device
with an environment, let us assume the total Hamiltonian can be
decomposed into a system part $H_\sys$, an environmental part
$H_\env$, and an interaction term $H_I$ (this presumes an
identification of some system of interest):
\begin{equation}
H = H_\sys \otimes \id_\env + \id_\sys \otimes H_\env + H_I,
\end{equation}
and that 
\begin{equation}
\left[ \ket{\varphi_j}\!\bra{\varphi_j} \otimes \id_\env , H \right] = 0 \quad \forall j.
\end{equation}
In such a case $\{ \ket{\varphi_j} \}$ is stable under the interaction with the environment, and is referred to as a \emph{pointer basis}. 
For simplicity we focus on exact commutation with the full Hamiltonian, but the conclusions that follow also hold for $\left[ \ket{\varphi_j}\!\bra{\varphi_j} \otimes \id_\env , H_I \right] = 0$  in the strong measurement limit, in which $H_I$ dominates the evolution~\cite{PazpointerbasisPRL1999}.

According to quantum theory the closed combined $\mathcal{S} + \mathcal{E}$ system evolves unitarily, with a state at time $t$ given by
\begin{align}
\label{eq:evolUnitary}
\rho_{\sys \env}(t) &= e^{-iHt/\hbar} \rho_{\sys \env}(0) e^{iHt/\hbar} \nonumber \\
&= \sum_{jk} \alpha_j \alpha^*_k \ket{\varphi_j}\!\bra{\varphi_k} \otimes \ket{E_j(t)}\!\bra{E_k(t)} ,
\end{align}
with the environmental states evolving according to  
\begin{equation}
\ket{E_j(t)} = e^{-i H_j t/\hbar} \ket{E_r}, 
\end{equation}
where $H_j \equiv \bra{\varphi_j} H \ket{\varphi_j}$ is the effective Hamiltonian that acts on the environment given that the system is in state $\ket{\varphi_j}$.

If the environment is large, correlations generated between system and environment typically cause rapid decay of the term $\bra{E_j(t)} E_k(t) \rangle$~\cite{schlosshauer2005review}. 
For many decoherence models this decay is exponential (see~\cite{schlosshauer2014review} and references within): 
\begin{equation}
\bra{E_j(t)} E_k(t) \rangle \approx \bra{E_j(0)} E_k(0) \rangle e^{-t/\tau_D}, \qquad j \neq k
\end{equation}
where the \emph{decoherence timescale} $\tau_D$ grows with the size of the bath.
In this way, the state of the system, $\rho_\sys(t)~=~\text{Tr}_\mathcal{E}\left[\rho(t)\right]$, given by
\begin{align}
\rho_{\sys}(t) &= \sum_{jk} \alpha_j \alpha^*_k \ket{\varphi_j}\!\bra{\varphi_k}  \bra{E_j(t)} E_k(t) \rangle \nonumber \\
&\approx \sum_{j} |\alpha_j|^2 \ket{\varphi_j}\!\bra{\varphi_j}\nonumber \\
&+ \sum_{j \neq k} \alpha_j \alpha^*_k \ket{\varphi_j}\!\bra{\varphi_k}  \bra{E_j(0)} E_k(0) \rangle e^{-t/\tau_D},
\end{align}
rapidly approaches the state that the system would be in if a collapse on the pointer basis had occurred:
\begin{align}
\rho_\sys^\ev(t) &= \sum_{j} |\alpha_j|^2 \ket{\varphi_j}\!\bra{\varphi_j}.
\end{align}
As mentioned in the introduction, the latter can be interpreted as a classical probability distribution. 
Hence, this dephasing process due to the interaction with an
environment implies that all measurements performed on the system will
give results exponentially close to those obtained on a system in
pointer states to which each of them can be assigned a classical probability.
It also provides an explanation for the inability to observe quantum mechanical effects except for tailored experimental situations, and effectively explains the quantum-to-classical transition~\cite{zurek1991,Zurek2003,riedel2010}.

This universal process is referred to as \emph{environmental decoherence}. For accessible introductions to the topic see~\cite{schlosshauer2005review,
schlosshauer2014review}. More thorough treatments can also be found in~\cite{schlosshauer2007book,joos2013}.

\section{No definite events in unitary quantum mechanics}
\label{sec:globalprotocol}

As we saw in the previous section, the effect of environmental
decoherence is that ---for all practical purposes--- 
it is as if the system has undergone a collapse after the interaction with $\env$, and as if an event has occurred.
However, given that the evolution of the total system is unitary, this is
not really the case, since the information that no event occurred is
still `there', hidden in environment--system correlations. 
 
This can be revealed by global system-environment protocols, an argument that can be famously traced back to Wigner~\cite{BookWigner1967}. Thereafter, 
d'Espagnat considered a concrete 
%$\sys+\env$ 
global observable on a particular spin-spin decoherence model for which the predictions differ depending on whether a definite event occurred or not~\cite{dEspagnat1995}.
More recently, Frauchiger and Renner built on Wigner friend's paradox, formally proving that in unitary quantum mechanics one cannot consistently define the occurrence of unique events in a self-consistent way~\cite{FrauchigerRenner2016}. 

In this section we review a simple protocol that can flesh out the fact that no event occurs in unitary quantum mechanics, related to that of Wigner, and of Frauchiger and Renner, and that unlike d'Espagnat's proposal works for any decoherence model. 

Concretely, consider that an extremely capable experimenter undoes the
evolution that the closed system $\sys+\env$ was subjected to.
That is, assume that a
global unitary $U$ is applied at time
$t$ %on $\mathcal{S} + \mathcal{E}$
that undoes the evolution given by Eq.~\eqref{eq:evolUnitary}.  Then,
the subsequent final state at a time $t_f$
is %posterior to the end of the protocol is
\begin{align}
\rho_{\sys \env}(t_f) &= U \rho_{\sys \env}(t_f) U^\dag \nonumber \\
&= \rho_\sys(0) \otimes \ket{E_r}\!\bra{E_r}.
\end{align}
Physically, this could \emph{in principle}  be implemented by a time reversal operation, which effectively evolves the system back in time $\{t\rightarrow -t\}$, so that $U = e^{i H t/\hbar}$. 
This sort of protocol, sometimes referred to as Loschmidt echo or spin echo~\cite{JalabertPastawskiPRL2001,PetitjeanPRL2006,LoschmidtEcho2012}, has been implemented in experiments, serving as a witness for chaoticicty~\cite{Peres84}.
More elaborate protocols have been proposed to implement such
operations even in cases without direct access to the system, and in
which the Hamiltonian is unknown~\cite{NavascuesPRX2018}. Notice that there are
limitations to reversing a system so in practice, there
are some errors in the protocols \cite{PetitjeanPRL2006}.
We are here assuming that the time reversal is perfect, but it is enough to assume that the experimenter is able to evolve backwards with a sufficient  precision such that it would allow to distinguish the two situations that we will consider here.

Clearly, the state $\rho_{\sys \env}(t_f)$, identical to the initial state, is not the state one would have if an event had occurred on the system during the evolution. In such a case, one would instead have
\begin{align}
\rho_{\sys \env}^\ev (t_f) &\equiv \sum_j \bra{\varphi_j} \rho_{\sys \env} (t_f) \ket{\varphi_j} \ket{\varphi_j}\! \bra{\varphi_j} \nonumber \\
&=  \sum_{j} |\alpha_j |^2 \ket{\varphi_j}\!\bra{\varphi_j} \otimes \ket{E_r}\!\bra{E_r} .
\end{align}
Note that, since $\{ \ket{\varphi_j}\}$ is a pointer basis which commutes with the Hamiltonian, the particular time at which the event occurs is irrelevant.

Indeed, simple local system observables of the form
\begin{equation}
O = O_\sys \otimes \id_\env
\end{equation}
are sufficient to distinguish between the states $\rho_{\sys \env}(t_f) $ and $\rho_{\sys \env}^\ev(t_f) $.  For example, for the observable $O_\sys~=~\ket{\varphi_j}\!\bra{\varphi_k}~+~\ket{\varphi_k}\!\bra{\varphi_j}$ one has
\begin{equation}
\label{eq:expectationOunitary}
\tr{O \rho_{\sys \env}(t_f) } = \alpha_j\alpha_k^* + \alpha_k\alpha_j^*,
\end{equation}
while
\begin{equation}
%\tr{O \rho_{\sys \env}^\ev(t_f) } = |\alpha_j|^2 + |\alpha_k|^2.
\tr{O \rho_{\sys \env}^\ev(t_f) } = 0.
\end{equation}

Implementing this protocol with enough approximation to distinguish
the two situations in an experiment is, without a doubt, an extremely
hard task, that would require control over the huge number of degrees
of freedom in the environment. However, knowing that this possibility
\emph{in principle} exists is already an insurmountable obstacle to
constructing an objective notion of event within unitary quantum
mechanics.  The fact that such an experiment in a macroscopic system
is way beyond our current technological capabilities is beside the
point: if the laws of physics in principle allow the protocol to work,
then an extremely skilled experimenter could in the future apply it to
the Earth and its surroundings, showing that the events around us were
a mere subjective experience.  While this sort of interpretation of
the universe may have
advantages~\cite{saunders2010many}%~\cite{Lev-manyworlds-encyclopedia}
, we are of the opinion that a realist interpretation with a clear cut
definition of events is much more preferable.

\section{Fundamental time and length uncertainties}
\label{sec:fundamentaluncertainties}

%The protocol considered in the previous section relies on unitary quantum mechanics.
%This is an implicit assumption in any physical process or experiment considered. In broad strokes one can decompose an arbitrary quantum mechanical process into preparation, evolution, and measurement. Since quantum mechanics is an intrinsically probabilistic theory, one in general needs a repetition of such process. 
%However, 

Being able to determine the time during which the initial unitary evolution given by Eq.~\eqref{eq:evolUnitary} takes places is crucial for the global protocol to work.
%In order for the global protocol to work, it is crucial to be able to accurately determine the time over which the initial unitary evolution takes place. 
Note that the unitary $U$ that inverts the dynamics has to be specifically tailored to counteract the evolution undergone during the time $t$.
In order to achieve this, one needs a physical system that can be used as a time tracking device, i.e. a \emph{clock}.

The limitations of quantum systems that can function as clocks have been extensively studied in the past~\cite{SaleckerWignerPhysRev1958,
PeresAmJPhys80,PeresBook2006,
AharonovPRA98,
MassarPRL99,booktimevol1,
booktimevol2, Khosla:2016tss, CastroRuizPNAS2017,
Hoehn1,Hoehn2,Hoehn3}, 
and has recently regained traction due to its relevance to the area of quantum thermodynamics~\cite{MalabarbaNJPhys2015,RankovicarXiv2015,
ErkerPRX2017,WoodsarXiv2018}.
A trait shared by quantum clocks is that quantum physics translates into uncertainties in the measurement of time. As is many times the case, these uncertainties decrease as the clock system becomes larger and/or more energetic. In fact, in most of these analyses the errors vanish in the limit of infinite energy. 
However, this limit clashes with general relativity, which imposes constraints on the energetic content allowed within a given region without significantly affecting physics around it.

Based on this argument, a series of authors have derived phenomenological limits on the measurements of time and length intervals on simple models~\cite{SaleckerWignerPhysRev1958,NgAnnals1995,
  LloydNature2000,BaezClassquantgrav2002,NgClassquantgrav2003, Frenkel:2010ai}

Here we adopt the smallest proposed errors,
% by combining quantum mechanical and general relativistic arguments 
%. In broad strokes the idea can be put as follows. Following XXX, 
\begin{subequations}
\label{eq:fundamentaluncertainties}
\begin{align}
\Delta_T &= T_P^{2/3} T^{1/3}, \qquad T_P \equiv \sqrt{\frac{\hbar G}{c^5}} \approx 5\times 10^{-44} s \\
\Delta_L &= L_P^{2/3} L^{1/3}, \qquad L_P \equiv \sqrt{\frac{\hbar G}{c^3}} \approx 2\times 10^{-35} m,
\end{align}
\end{subequations}
where $T_P$ and $L_P$ are Planck time and length respectively. Notice the \emph{tiny} errors that these entail: $\Delta T \sim 10^{-23} s$ on a timescale of the age our galaxy, and $\Delta L \sim 10^{-16} m$ over its lengthscale. 
%We also note that the precise exponents are not particularly relevant for our following arguments, and corrections would lead to similar conclusions.
 %, $c$ is the speed of light in vacuum and $G$ is the gravitational constant.

\begin{figure}
\includegraphics[width= 1\columnwidth]{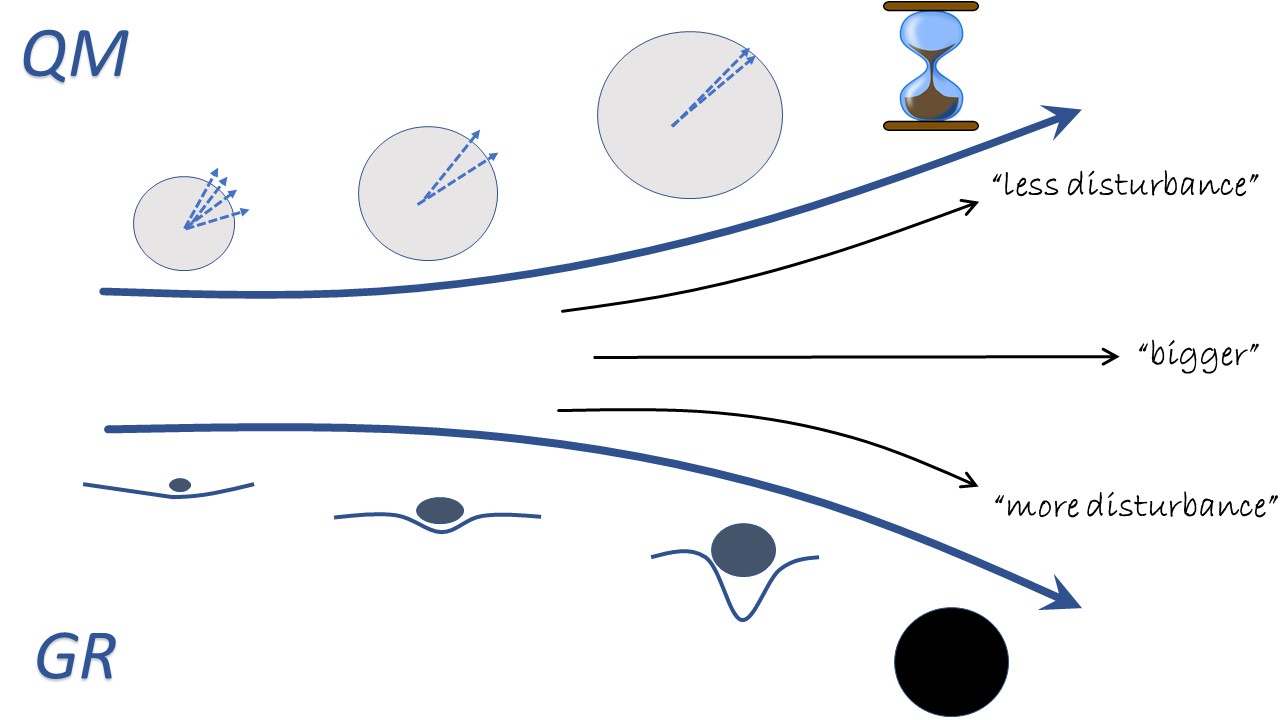}
\caption{\label{fig:GRQM} 
 {\bf Fundamental limits on clocks.} 
Detailed studies on models for quantum clocks indicate that, as is typically the case in quantum systems, relative uncertainties decrease with the energy of the system.
However, increasing the energetic content comes at a cost, since general relativity effects start to play a role. 
Ultimately, the maximum energy that can be stored within a region is bounded, given that once the system becomes a black hole no information can be retrieved, and the system becomes useless as a clock. Phenomenological analyses combining these arguments have been used to derive fundamental limits on the accuracy with which time and length intervals can be measured.
}
\end{figure}

We ultimately take these fundamental uncertainties as an assumption at this stage, grounded on the concept that ideal time and length measurements are unlikely to exist when combining general quantum mechanical and general relativistic principles. Such uncertainties would limit the accuracy of any device used as clock, as illustrated on Fig.~\ref{fig:GRQM}.
Admittedly, the heuristic arguments used to derive $\Delta_T$ and $\Delta_L$ might not end up being exact in the light of a theory combining quantum mechanics and general relativity. 
Nevertheless, while we 
use~\eqref{eq:fundamentaluncertainties} in the remainder of the paper, we note that the exact scaling with $T$ and $L$ will not be particularly relevant for the qualitative subsequent conclusions.

%The inability to perfectly access 

These fundamental uncertainties influence physics. Indeed, since the ideal parameter $t$ is inaccessible, it is unphysical to express the evolution of what can be observed of a system in terms of it. 
Instead, the physically meaningful approach is to express physics relationally, solely in terms of observable quantities. 

In order to model this we consider a physical device used to track
time. i.e. a clock, with a dynamics dictated by a Hamiltonian
$H_\mathcal{C}$, that for simplicity we assume evolves independently
from the system of interest (in this case $\sys +\env$) with a
Hamiltonian $H$.  Then, the full Hamiltonian of
$\sys + \env + \mathcal{C}$ is
\begin{equation}
H_{\textnormal{total}} = H\otimes \id_\mathcal{C} + \id_{\sys \env} \otimes H_\mathcal{C}.
\end{equation}
Hence, given an initially uncorrelated clock the full state evolves according to
\begin{align}
\rho_{\textnormal{total}}(t) &= e^{-i H_{\textnormal{total}} t /\hbar} \rho_{\sys \env}(0)\otimes\rho_{\mathcal{C}}(0) e^{i H_{\textnormal{total}} t /\hbar} \nonumber \\
&= e^{-i H t /\hbar} \rho_{\sys \env}(0) e^{i H t /\hbar}\otimes e^{-i H_{\mathcal C} t /\hbar} \rho_{\mathcal{C}}(0) e^{i H_{\mathcal C} t/\hbar } \nonumber \\
& = \rho_{\sys \env}(t)\otimes\rho_{\mathcal{C}}(t).
\end{align}

In order for the clock to act as such, some observable 
\begin{align}
\hat T = \sum_T T \hat{\Pi}_T
\end{align}
has to be a reliable time-tracking observable, where $\hat{\Pi}_T$ is a projector onto the subspace corresponding to eigenvalue $T$ of $\hat T$. That is, the evolution of $\tr{\hat T \rho_{\mathcal C}(t)}$ should track the ideal time $t$. 
Then, a meaningful way to describe the dynamics of an observable $\hat
O = \sum_O O \hat{\Pi}_O$ is to consider  the conditional probability
of observing a particular value $O$ given that the \emph{physical
  time} takes a value $T$.
Such conditional probability is given by
\begin{align}
\mathcal{P}(O \vert T ) = \lim_{t_0\to
  \infty}\frac{\displaystyle\int_{-t_0}^{t_0}\text{Tr} \Big[ P_O P_T
  \rho_{\sys \env}(t) P_T \Big] dt}{ \displaystyle\int_{-t_0}^{t_0} \text{Tr} \Big[ P_T \rho_{\sys \env}(t) \Big] dt},
\end{align}
where the unobservable parameter $t$ is integrated
over~\cite{Gambini2007pedagogical,Gambini:2008ke}.

 The approach followed here is
  naturally geared towards a relational description of the world, like
  the one provided by general relativity. Our approach would work even
  in a description where time is an emergent property of the universe.
  Page and Wootters \cite{Page:1983uco} pioneered the use of a relational time in general
  relativity and some shortcomings of their approach were addressed in
  \cite{Gambini:2008ke}.

Interestingly, the traditional rule for calculating probabilities is recovered,
\begin{align}
\mathcal{P}(O \vert T ) &= \text{Tr} \bigg( P_O \int  \rho_{\sys \env}(t) \mathcal{P}_t(T) dt \bigg)  \nonumber \\
&\equiv \text{Tr} \big( P_O \rho_{\sys \env}(T) \big),
\end{align} 
where with some abuse of notation we denote
\begin{align}
\rho_{\sys \env}(T) \equiv \int  \rho_{\sys \env}(t) \mathcal{P}_t(T) dt,
\end{align}
and
\begin{align}
\mathcal{P}_t(T) \equiv \frac{ \text{Tr} \big[P_T \rho_{\mathcal C}(t)
  \big] }{\lim_{t_0\to \infty}\int_{-t_0}^{t_0} \text{Tr} \big[ P_T \rho_{\mathcal C}(t) \big] dt}
\end{align} 
is the probability of getting $T$ at the ideal time $t$.
The \emph{effective state} $\rho_{\sys \env}(T)$ then gives a characterization of the observable physics in terms of the physical time $T$.

With access to an ideal clock, for which ${P}_t(T) = \delta(T-t)$, one immediately recovers the traditional quantum mechanical result. 
However, fundamental time uncertainties lead to non-unitary evolution. 
Here is a simple way to illustrate it. Consider, for simplicity, that the clock is characterized by a Gaussian distribution centered around the ideal time value $t$, with a standard deviation given by the fundamental uncertainty on measuring time intervals.
That is, assume
 \begin{align}
\label{eq:gaussianclock} 
 \mathcal{P}_t(T) = \frac{1}{\sqrt{2 \pi \Delta_T^2}} e^{-\frac{(T-t)^2}{2 \Delta_T^2}},
 \end{align}
where $\Delta_T = T_P^{2/3} T^{1/3}$. 
Expressing the initial density matrix in the energy eigenbasis as
\begin{equation}
\rho_{\sys \env}(0) = \sum_{nm} \rho_{nm} \ket{n} \bra{m}, \qquad H = \sum_n E_n \ket{n}\bra{n},
\end{equation}
the evolution in terms of ideal time $t$ is given by
\begin{align}
\rho_{\sys \env}(t) = \sum_{nm} \rho_{nm} e^{-i(E_n - E_m) t/\hbar} \ket{n} \bra{m}.
\end{align}
From here the evolution in terms of a physical time then becomes
\begin{eqnarray}
\rho_{\sys \env}(T) &=& \sum_{nm} \rho_{nm} e^{-i(E_n - E_m)
                        T/\hbar}\nonumber\\
  &&\times e^{-(E_n - E_m)^2 \Delta_T^2/(2\hbar^2)} \ket{n} \bra{m}.
\end{eqnarray}
Note that it can be decomposed into unitary and non-unitary parts:
\begin{equation}
\label{eq:dephasingplusunitary}
\rho_{\sys \env}(T) = e^{-iHT/\hbar} \Omega_T(\rho_{\sys \env}(0)) e^{iHT/\hbar}, 
\end{equation}
where the map $\Omega_T$ is defined by
\begin{equation}
\Omega_T(\ket{n}\bra{m}\!) = \sum_{nm} e^{-(E_n - E_m)^2 \Delta_T^2/(2\hbar^2)} \ket{n} \bra{m}.
\end{equation}

The physically accessible density matrix $\rho_{\sys \env}(T)$ then evolves according to the master equation
\begin{eqnarray}
\label{eq:mastereqphysical}
\frac{\partial \rho_{\sys \env}(T)}{\partial T} &=& -\frac{i}{\hbar}
                                                    [H,\rho_{\sys
                                                    \env}(T)]\nonumber\\
  &&- \frac{1}{2\hbar^2} \frac{ \partial (\Delta_T^2)}{ \partial T}
     \left[ H, [H, \rho_{\sys \env}(T)]\right].
\end{eqnarray}
While the first term corresponds to unitary quantum theory, the second term induces a loss of coherence in the energy basis, strictly due to the fundamental time uncertainties. 
In~\cite{Gambini2007pedagogical}
it is shown that for short times the effective state undergoes such non-unitary dynamics without the assumption of a Gaussian distribution for the clock.
Note that similar evolutions have been considered in the literature, with different motivations~\cite{MilburnPRA1991,
EgusquizaPRA1999,diosi2005intrinsic}. Here the crucial point is that,
under the assumption that ideal time measurements are forbidden in
nature, such non-unitary evolution is \emph{fundamental}. Any system
described relationally is inevitably bound by these limits, and its
evolutions will be described by
the master equation~\eqref{eq:mastereqphysical}.
Note, too, that the above analysis can be extended to include decoherence due to errors in measuring length scales in quantum field theory. As such, even if we present our results within non-relativistic quantum mechanics, they are naturally geared towards covariant formulations of quantum theory~\cite{GPPIntJModPhys2006}.

A question that naturally arises at this point is whether enlarging the analysis to include the clock system could change the results, i.e. how fundamental is this process, when it depends on the clock system? 
While our description explicitly includes a physical device used as clock, such loss of coherence is inescapable. Imagine the set of all physical processes in the universe that can serve to track time. If fundamental uncertainties exist, all such processes will be restricted by them. Then, the best one could aim for when describing another physical process is having such optimal clock in hand, as we have done above. Any realistic setting, with a realistic clock, will at least suffer from the above loss of coherence.

\section{Production of events}
\label{sec:events}

Let us analyze the global protocol introduced in Sec.~\ref{sec:globalprotocol} in terms of the fundamentally non-unitary evolution of the total closed system $\sys + \env$ due to formulating physics in terms of a measurable time $T$.
%The evolution of $\mathcal{S} + \mathcal{E}$ in terms of a physical time variable $T$ is easy to express in the eigenbasis of the Hamiltonian:
%\begin{align}
%\rho(T) = \sum_{nm} \rho_{nm} e^{-i(E_n - E_m) T} e^{-(E_n - E_m)^2 \Delta_T^2/2} \ket{n} \bra{m},
%\end{align}
%where $H = \sum_n E_n \ket{n} \bra{n}$.
%There is a fundamental energy decoherence during the initial time $t$ of the protocol, when the system evolves according to Hamiltonian $H$, as well as during the final time $t$, when the opposite Hamiltonian $-H$ is driving the state towards $\rho(0)$.
First, note that in the best case scenario the global unitary $U$ undoes the unitary part of the evolution in Eq.~\eqref{eq:dephasingplusunitary} given a previous timespan $T$.
Then, the final state expressed in terms of a physical time $T$ is
%\begin{align}
%\rho_{\sys \env}(T_f) = \sum_{nm} \bra{n} \rho_{\sys \env}(0) \ket{m} e^{-(E_n - E_m)^2 \Delta_T^2/2} \ket{n} \bra{m},
%\end{align}
\begin{align}
\rho_{\sys \env}(T_f) = \sum_{nm}  \rho_{nm}  e^{-(E_n - E_m)^2 \Delta_{2T}^2/(2\hbar^2)} \ket{n} \bra{m},
\end{align}
where we are assuming implementation of the unitary $U$ by a time-reversal operation during a time $T$, resulting in a total time $2T$ over which loss of coherence occurs.

%[ XXX PUSIMOS UN  FACTOR DOS POR IDA Y VUELTA. ESTAS DE ACUERDO? HABRIA QUE
%INCLUIRLO EN LAS SUBSIGUIENTES XXX]
%where $\Delta_T^2 = T_P^{4/3} T^{2/3}$ is due to the fundamental decoherence over a time $T$.  
%%If the implementation of $U$ were attempted 
%\textcolor{blue}
%{
%We are assuming the best case scenario of perfect implementation of the unitary $U$. If this unitary were implemented by a time-reversal operation during a time $T$, then there would be an extra decoherence, over the total duration of the unitary evolution of $2T$.  
%}

Using the fact that $e^{-(E_n - E_m)^2 \Delta_{2T}^2/(2\hbar^2)} = \sqrt{\frac{1}{ 2 \pi \Delta_T^2}} \int_{-\infty}^{\infty} e^{-\mu^2 /(2\Delta_{2T}^2)} e^{-i(E_n -E_m) \mu/\hbar} d\mu$, the final state can be expressed as
%\begin{align}
%\rho_{\sys \env}(T_f)  &= \sum_{nm} \bra{n} \rho_{\sys \env}(0)  \ket{m} \sqrt{\frac{2}{\pi \Delta_T^2}} \nonumber \\
%& \int_{-\infty}^{\infty} e^{-2 \mu^2 /\Delta_T^2} e^{-i(E_n -E_m) \mu} d\mu \ket{n} \bra{m} \nonumber \\
%&= \sqrt{\frac{2}{\pi \Delta_T^2}} \int_{-\infty}^{\infty} e^{-2 \mu^2 /\Delta_T^2} \widetilde{\rho}_{\sys \env}(\mu) d\mu,
%\end{align}
\begin{align}
\rho_{\sys \env}(T_f)  &= \sum_{nm}  \rho_{nm}  \sqrt{\frac{1}{2 \pi \Delta_{2T}^2}} \nonumber \\
& \int_{-\infty}^{\infty} e^{- \mu^2 /(2\Delta_{2T}^2)} e^{-i(E_n -E_m) \mu/\hbar} d\mu \ket{n} \bra{m} \nonumber \\
&= \sqrt{\frac{1}{ 2 \pi \Delta_{2T}^2}} \int_{-\infty}^{\infty} e^{-\mu^2 /(2\Delta_{2T}^2)} \widetilde{\rho}_{\sys \env}(\mu) d\mu,
\end{align}
where $\widetilde{\rho}_{\sys \env}(\mu) \equiv e^{-i H \mu} \rho_{\sys \env}(0) e^{i H \mu}$ is the unitarily evolved state (in order to avoid confusion in this section we distinguish it from the effective state at a physical time $T$ by a tilde). 
Similarly, 
\begin{align}
\rho_{\sys \env}^\ev(T_f) &= \sqrt{\frac{1}{2 \pi \Delta_{2T}^2}} \int_{-\infty}^{\infty} e^{-\mu^2 /(2 \Delta_{2T}^2)} \widetilde{\rho}_{\sys \env}{}^\ev(\mu) d\mu.
\end{align}

The unitary evolution of system plus environment was analyzed in Section~\ref{sec:environmentaldecoherence}.
From Eq.~\eqref{eq:evolUnitary} we obtain that for an observable $O = O_\sys \otimes \id_\env$:
\begin{align}
&\tr{O \big( \rho_{\sys \env}(T_f)  -\rho_{\sys \env}^\ev(T_f)  \big) } \nonumber \\
&=
 \sqrt{\frac{1}{2 \pi \Delta_{2T}^2}} \int_{-\infty}^{\infty} e^{- \mu^2 /(2\Delta_{2T}^2)}
 \tr{O \big( \widetilde{\rho}_{\sys \env}(\mu)  - \widetilde{\rho}_{\sys \env}^\ev(\mu)  \big) }  d\mu\nonumber \\
&=
 \sqrt{\frac{1}{2 \pi \Delta_{2T}^2}} \int_{-\infty}^{\infty} e^{- \mu^2 /(2\Delta_{2T}^2)} d\mu
 \nonumber \\
 & \qquad \sum_{j \neq k} \alpha_j \alpha_k^* \tr{O_S \ket{\varphi_j}\!\bra{\varphi_k} } \bra{E_j(\mu)} E_k(\mu) \rangle.
\end{align}
For a great number of decoherence models the overlap between the environmental states decays exponentially~\cite{schlosshauer2014review}, 
%$\bra{E_j(\mu)} E_k(\mu) \rangle \approx e^{-|\mu|/\tau_D} \bra{E_j(0)} E_k(0) \rangle$, so
in which case
\begin{align}
&\tr{O \big( \rho_{\sys \env}(T_f)  -\rho_{\sys \env}^\ev(T_f)  \big) } \approx \nonumber \\
& \qquad \sqrt{\frac{1}{2\pi \Delta_{2T}^2}} \int_{-\infty}^{\infty} e^{-  \mu^2 /(2\Delta_{2T}^2)} e^{-|\mu|/\tau_D} d\mu \nonumber \\
& \qquad \quad \qquad \sum_{j \neq k} \alpha_j \alpha_k^* \tr{O_S \ket{\varphi_j}\!\bra{\varphi_k} } \bra{E_j(0)} E_k(0) \rangle \nonumber \\
 & = e^{\Delta_{2T}^2/2\tau_D^2} \, \text{erfc}\left(\frac{\Delta_{2T}}{\sqrt{2}\tau_D} \right)    \tr{O \big( \rho_{\sys \env}(0)  -\rho_{\sys \env}^\ev(0)  \big) } \nonumber \\
 & \qquad  \leq \frac{\sqrt{2}\tau_D}{\sqrt{\pi}\Delta_{2T}}     \tr{O_\sys \big( \rho_{\sys}(0)  -\rho_{\sys}^\ev(0)  \big) },
\end{align}
where we used in the last step that the environment and system are not correlated initially.

Given Eqs.~\eqref{eq:fundamentaluncertainties} as an estimate for the error in the physical time, we obtain
\begin{align}
\label{eq:observabledecay}
&\tr{O \big( \rho_{\sys \env}(T_f)  -\rho_{\sys \env}^\ev(T_f)  \big) } \nonumber \\
&\qquad \lesssim \frac{\sqrt{2}\tau_D}{\sqrt{\pi} 2^{1/3} T_P^{2/3}T^{1/3}}     \tr{O_\sys \big( \rho_{\sys }(0)  -\rho_{\sys  }^\ev(0)  \big) }.
\end{align}
This shows that using the global protocol to distinguish the evolved state from the state in case of an event becomes increasingly harder when time uncertainties are taken into account. 
The state thus becomes physically increasingly similar to the case in which an event occurs and it can be interpreted as a classical mixture.

\section{A fundamental criterion for the production of events}
\label{sec:definiteevents}

While distinguishing between $\rho_{\sys \env}(0)$ and $\rho_{\sys \env}^\ev(0)$ becomes increasingly hard when taking into account the loss of coherence due to uncertainties in time measurements, this is still a solution that works `for all practical purposes' at this stage, given that in principle extremely precise measurements could distinguish them.
We now show that {\em this is not the case}  when one takes into account uncertainties on length measurements. We illustrate it in the paradigmatic case of a particle in a coherent superposition over two spatial locations.

For concreteness, consider that the initial state of the system is a 1-D coherent spatial superposition of Gaussian wavepackets of width $\sigma$ separated by a distance $L$:
% centered around positions $x_1$ and $x_2$:
\begin{align}
\rho_\sys(0) &= |a|^2 \ket{\varphi_1}\!\bra{\varphi_1} + |b|^2 \ket{\varphi_2}\!\bra{\varphi_2} \nonumber \\
& + ab^* \ket{\varphi_1}\!\bra{\varphi_2} + a^*b \ket{\varphi_2}\!\bra{\varphi_1},
\end{align}
with
\begin{align}
\langle x \ket{\varphi_1} &= \frac{1}{(2\pi\sigma^2)^{1/4}} \exp\left(-\frac{(x-L/2)^2}{4\sigma^2}\right) \\
\langle x \ket{\varphi_2} &= \frac{1}{(2\pi\sigma^2)^{1/4}} \exp\left(-\frac{(x+L/2)^2}{4\sigma^2}\right).
\end{align}
The corresponding state given the occurrence of an event on position is 
\begin{align}
\rho_\sys^\ev(0) &= |a|^2 \ket{\varphi_1}\!\bra{\varphi_1} + |b|^2 \ket{\varphi_2}\!\bra{\varphi_2}.
\end{align}
(Note that, while $\ket{\varphi_1}$ and $\ket{\varphi_2}$ are not orthonormal, their overlap is small as long as $L \gg \sigma$.)

%\begin{align}
%\langle x \ket{\varphi_j} = \frac{1}{(2\pi\sigma^2)^{1/4}} \exp\left(-\frac{(x-x_j)^2}{4\sigma^2}\right), \quad j = \{1,2 \}.
%\end{align}

Choosing the momentum operator $O_S = p$ can serve to distinguish between a coherent superposition and a statistical mixture over different spatial positions. Indeed, given that
%\begin{align}
%&\bra{\varphi_j} P \ket{\varphi_k} = \nonumber \\
%& \frac{-i \hbar}{\sqrt{2\pi\sigma^2}}\int \frac{-2(x-x_k)}{4\sigma^2}\exp\left(-\frac{(x-x_j)^2+(x-x_k)^2}{4\sigma^2}\right)dx \nonumber \\
%&\qquad = \frac{2i \hbar}{\sqrt{2\pi\sigma^2}}\int  u \exp\left(-u^2 -\left( u+\frac{x_k-x_j}{2\sigma}\right)^2\right) du\nonumber \\
%&\qquad = -  i \hbar \frac{\left( x_k-x_j \right)}{4\sigma^2}  \exp\left(-\frac{\left(x_k-x_j\right)^2}{8\sigma^2} \right) , 
%\end{align}
\begin{align}
\bra{\varphi_1} p \ket{\varphi_1} = \bra{\varphi_2} p \ket{\varphi_2} = 0,  
\end{align}
we have 
\begin{align}
\tr{p \rho_\sys^\ev(0)} = 0.
\end{align}
Meanwhile, using that 
\begin{align}
&\bra{\varphi_2} p \ket{\varphi_1} = \nonumber \\
 & \frac{-i \hbar}{\sqrt{2\pi\sigma^2}}\int \frac{-2(x-L/2)}{4\sigma^2}\nonumber\\
&\times    \exp\left(-\frac{(x+L/2)^2+(x-L/2)^2}{4\sigma^2}\right)dx \nonumber \\
&\qquad = \frac{2i \hbar}{\sqrt{2\pi\sigma^2}}\int  u \exp\left(-u^2 -\left( u+\frac{L}{2\sigma}\right)^2\right) du\nonumber \\
&\qquad = -  i \hbar \frac{  L  }{4\sigma^2}  \exp\left(-\frac{L^2}{8\sigma^2} \right) , 
\end{align}
gives
%if the distance between the two lobes is denoted by $L \equiv x_1 - x_2$:
\begin{align}
\tr{p \rho_\sys(0)} = - i(ab^* - a^*b) \hbar \frac{L}{4\sigma^2}  \exp\left(-\frac{L^2}{8\sigma^2} \right).
\end{align}
Hence, if the initial state is chosen with the appropriate phases, $p$ discriminates whether an event occurred or not.

Note, however, that %an uncertainty $\Delta L$ on the distance $L$ between the two lobes translates into an error on the expectation value of $P$. A simple propagation of uncertainty gives
fundamental uncertainties on the measurement of length intervals 
forbid a perfect preparation of the wavepackets $\ket{\varphi_1}$ and $\ket{\varphi_2}$, since they imply errors 
 $\Delta L$ and $\Delta \sigma$ on the separation and width of the lobes.
This in turn
translates into uncertainties on the expectation value of $p$.
A simple propagation of uncertainty on the error induced in $\tr{p \rho_\sys(0)}~-~\tr{p \rho_\sys^\ev(0)}$ gives
\begin{align}
%\label{eq:uncertaintyobservable}
 &\Delta^2 \left( \tr{p \rho_\sys(0)} - \tr{p \rho_\sys^\ev(0)} \right) & \nonumber \\
 &\quad  = \left( \frac{\partial }{\partial L}  \tr{p \rho_\sys(0)} \right)^2 \Delta L^2 + \left( \frac{\partial }{\partial \sigma}  \tr{p \rho_\sys(0)} \right)^2 \Delta \sigma^2 \nonumber \\
 &\quad = \left( \left( \frac{1}{ L} - \frac{ L }{ 4\sigma^2 } \right)\tr{p \rho_\sys(0)}  \right)^2 \Delta L^2 \nonumber \\
 &\quad + \left( \left( - \frac{2}{ \sigma } + \frac{ L^2 }{ 4\sigma^3} \right)\tr{p \rho_\sys(0)}  \right)^2 \Delta \sigma^2. 
\end{align}
Expressing the standard deviation of the Gaussians as $\sigma = \epsilon L$, we get
\begin{align}
\label{eq:uncertaintyobservable}
 &\Delta^2 \left( \tr{p \rho_\sys(0)} - \tr{p \rho_\sys^\ev(0)} \right) & \nonumber \\
 &\quad = \left(  1 - \frac{ 1 }{ 4\epsilon^2 }  \right)^2 \tr{p \rho_\sys(0)}^2 \frac{\Delta L^2}{L^2} \nonumber \\
 &\quad + \left(  - \frac{2}{ \epsilon } + \frac{ 1 }{ 4\epsilon^3 } \right)^2 \tr{p \rho_\sys(0)}^2 \epsilon^2 \frac{\Delta L^2}{L^2} \nonumber \\
 &\quad = \left(  5 - \frac{ 1 }{ \epsilon } - \frac{ 7 }{ 16\epsilon^2 } + \frac{ 1 }{ 16\epsilon^4 }  \right)  \tr{p \rho_\sys(0)}^2 \frac{\Delta L^2}{L^2} \nonumber \\
 &\quad \ge 2  \tr{p \rho_\sys(0)}^2 \frac{\Delta L^2}{L^2},
\end{align}
with the inequality valid for any choice of $\epsilon \ge 0$.

\begin{figure*}
\includegraphics[width= 1.75\columnwidth]{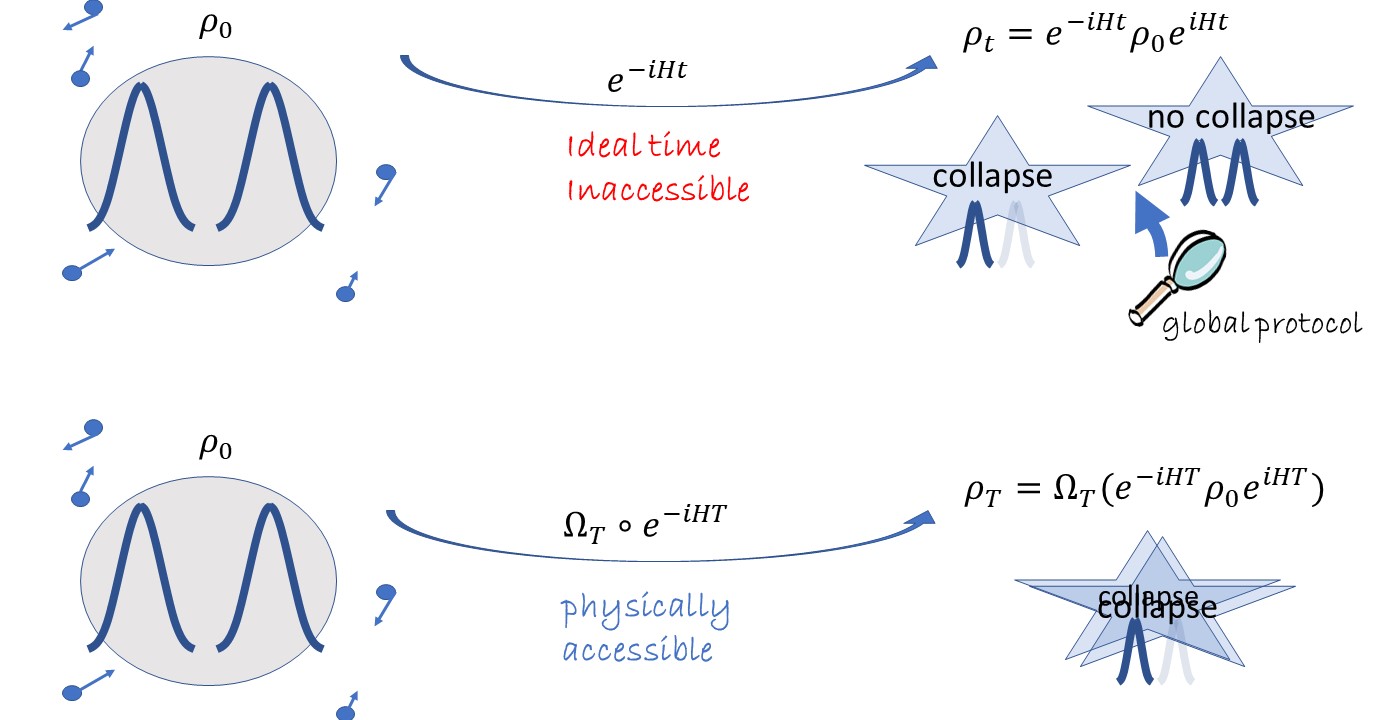}
\caption{
  \label{fig:events} {\bf Production of events.}  (Top) Within
  ordinary unitary quantum theory, there exists global protocols,
  i.e. unitary operations followed by measurements, that allow to
  distinguish the unitarily evolved state from the state one would
  have if an event, or collapse, had occurred.  Thus, the final state
  cannot be interpreted as a classical probability distribution over
  the set of pointer states. In the example shown, a definite position
  for the particle cannot be assigned.  (Bottom) However, when the
  loss of coherence due to uncertainties in time measurements is taken
  into consideration, the power of these global protocols to
  distinguish the two situations is diminished. After a long enough
  period of time has passed, fundamental uncertainties in the initial
  state of the system, in this example related to the initial position
  of the particle, forbid one from physically distinguishing the
  evolved state from a state where the particle ends up in a definite
  position, with a certain probability.  In such situations, when the
  state is physically completely indistinguishable from a classically
  interpretable state (statistical mixture), we say that an event has occurred.  }
\end{figure*}

The uncertainty on the measurement of $O = p \otimes \id_\env$ has to be taken into account when analyzing the global protocol considered in the previous section. 
Once the condition
\begin{align}
&\tr{p \big( \rho_{\sys \env}(T_f)  -\rho_{\sys \env}^\ev(T_f)  \big) } \nonumber \\
&\qquad \quad \le \Delta \left( \tr{p \left(\rho_\sys(0) - \rho_\sys^\ev(0) \right)} \right) 
\end{align}
is satisfied, 
the uncertainty in the measurement of the observable prevents one from
verifying whether the system is in a coherent superposition
$\rho_{\sys \env}(0)$ or in a statistical mixture $\rho_{\sys
  \env}^\ev(0)$, as illustrated in Fig.~\ref{fig:events}.
Given the last bound on Eq.~\eqref{eq:uncertaintyobservable}, this eventually happens for large enough $T_f$.
Notice that
this is a fundamental limitation and cannot be circumvented by making
multiple measurements. It is related to the impossibility of preparing
exactly the same initial state with infinite precision. This is an
explicit implementation of the notion of undecidability between a
mixture state and one resulting from the evolution that, with an
example of a spin system, was used in formulating the Montevideo
Interpretation of quantum mechanics~\cite{GGPPundecidability2010,GGPPcriterionevents2011}.

In order to estimate how fast one arrives at the condition described
by equation (47),  we focus on a particular decoherence model. 
In the past our attention went to models of decoherence on a spin
degree of freedom, so for completeness we now analyze decoherence of
spatial superpositions due to a scattering processes with an environment (see for example Chapter 3 of~\cite{schlosshauer2007book} for a detailed presentation).
Adopting the terminology of that presentation, for a ``central system'' $\sys$ of mass $M$ and cross-section $a$ undergoing scattering events with a bath of smaller particles,  a spatial superposition over a distance $L$ exponentially decays on a timescale
\begin{equation}
\label{eq:timedecoh}
\tau_D = \frac{1}{ \Lambda L^2},
\end{equation}
where 
\begin{equation}
\label{eq:scattering}
\Lambda = \frac{\sqrt{2\pi}8}{3\hbar^2} \nu \sqrt{M} a^2 (K_B \text{Temp})^{3/2}
\end{equation}
is the scattering constant, and depends on characteristics of central system, the environment, and the interaction between them. Here, %$M$ is the mass of the central system,
 $\nu$ is the density of particles in the environment of temperature $\text{Temp}$, and $K_B$ is Boltzmann's constant. 
Schlosshauer gives a detailed treatment of the decoherence timescales
of different central systems, and in the presence of a variety of
baths. For instance, an atomic-sized dust grain ($a \sim 10^{-8}m $) initially in a superposition over $L =a$ is decohered by air at room temperature and normal pressure ($\nu \approx 3\times10^{23} m^{-3}$) on a timescale $\tau_D \sim 10^{-31}s$. 

For illustration purposes we follow Schlosshauer, and take superpositions over distances of the order of the size of the central system, $L = a$.
Combining Eqs.~\eqref{eq:fundamentaluncertainties},~\eqref{eq:observabledecay} and~\eqref{eq:uncertaintyobservable}, we conclude that an event occurs for times $T > \tau_\ev$, where the \emph{event time} $\tau_\ev$ satisfies
%\begin{align}
% \frac{\sqrt{2}\tau_D}{\sqrt{\pi}T_P^{2/3}\tau_\ev^{1/3}} = \frac{ 5}{2L} \Delta L = \frac{5 L_P^{2/3}}{2L^{2/3}}.
%\end{align}
\begin{align}
 \frac{\sqrt{2}\tau_D}{\sqrt{\pi} 2^{1/3} T_P^{2/3}\tau_\ev^{1/3}} = \frac{\sqrt{2} \Delta L}{L}  = \frac{ \sqrt{2}  L_P^{2/3}}{L^{2/3}}.
\end{align}
That is, for times $T > \tau_\ev$, fundamental uncertainties on time and length intervals prevent one from distinguishing $\rho_{\sys \env}$ from $\rho_{\sys \env}^\ev$ with global protocols, where
%\begin{align}
%\label{eq:timeevent}
%\tau_\ev =  \frac{2^4 \sqrt{2} }{ 5^3 \pi^{3/2}  } \frac{ \tau_D^3 L^2}{ T_P^2 L_P^2}.
%\end{align} 
\begin{align}
\label{eq:timeevent}
\tau_\ev =  \frac{ 1 }{ 2 (2\pi)^{3/2}  } \frac{ \tau_D^3 L^2}{ T_P^2 L_P^2}.
\end{align} 
From then on, physical predictions are exactly as if the system is found in a statistical mixture.
%
%The event timescale can be evaluated for different central systems. 
%For a dust grain this is still an extremely large timescale, $\tau_\ev \sim 3\times10^{45} s$, i.e. no event occurs within the lifetime of the universe.
%However, for a particle of the same mass density as a dust grain and characteristic size $\sigma \sim 10^{-4} m$ the estimate of the event timescale becomes XXXXXXXXXXXXXX $\tau_\ev \sim 1 s$ XXXXXXXXXXXXXX, and much smaller timescales are obtained for larger systems (this is readily derived from eqs.~\eqref{eq:timedecoh},~\eqref{eq:scattering} and \eqref{eq:timeevent}, since scaling the characteristic length of the particle by a factor $\alpha$ decreases the event time by a factor $\alpha^{-23/2}$). 
%Remarkably, the quantum/classical boundary is set on the verge of the macroscopic scale, from this simple analysis. 
%Naturally, a more realistic analysis could shift this line, as would a change in the fundamental time and length uncertainties. However, given the dependence of $\tau_\ev$ on the environmental decoherence timescales $\tau_D$, and the remarkable short values of the latter for big systems, events will invariably occur for macroscopic systems.

The event timescale can be evaluated for different central systems.
For an atomic-sized dust grain ($a \sim 10^{-8}m $) this is still an
extremely large timescale, $\tau_\ev \sim 1\times10^{44} s$, i.e. no
event occurs within the lifetime of the universe.  However, for a
larger particle of the same mass density as a dust grain and
characteristic size $a \sim 10^{-5} m$ the estimate of the event
timescale becomes $\tau_\ev \sim 50 s$, while for one of
characteristic size $a \sim 10^{-4} m$ events happen extremely
quickly, within $\tau_\ev \sim 10^{-12}s$.
% XXXXXXXXXXXXXX, and much smaller timescales are obtained for larger systems 
Such estimates are readily derived from eqs.~\eqref{eq:timedecoh},~\eqref{eq:scattering} and \eqref{eq:timeevent} by noting that scaling the characteristic length of the particle by a factor $\alpha$ decreases the event time by a factor $\alpha^{-29/2}$. 
Remarkably, the quantum/classical boundary is set on the verge of the macroscopic scale, from this simple analysis. 
Naturally, a more realistic analysis could shift this line, as would a change in the fundamental time and length uncertainties. However, given the dependence of $\tau_\ev$ on the environmental decoherence timescales $\tau_D$, and the remarkable short values of the latter for big systems, events will invariably occur for macroscopic systems. 
Note that detailed previous analysis on spin-spin decoherence models show that more stringent timescales for events can occur in other cases~\cite{GGPPundecidability2010,
GGPPcriterionevents2011}.

\section{A consistent single-world interpretation of quantum mechanics}
\label{sec:frauchigerrenner}

 Frauchiger and Renner recently showed that no ``single-world”
 interpretation of quantum mechanics can be self-consistent~\cite{FrauchigerRenner2016} 
 (for accessible colloquial and pedagogical presentations see~\cite{pusey2018inconsistent,BlogMateus,delRio2018arXiv}). 
 That is, if quantum mechanics exactly describes complex systems
 like the observers and their measuring devices, one needs to either give up
 ``the view that there is a single reality'', or the expectation of consistency between observations of different agents. A
 single-world interpretation is any interpretation that asserts, for a
 quantum measurement with multiple possible outcomes, that just one
 outcome and its corresponding reduction of the state actually occurs.
 
%\textcolor{blue}
%{
%Frauchiger and Renner recently showed that no
% interpretation of quantum mechanics where agents observe single outcomes in experiments can be self-consistent~\cite{FrauchigerRenner2016} 
% (for accessible colloquial and pedagogical presentations see~\cite{BlogMateus} and~\cite{delRio2018arXiv}). 
% That is, if quantum mechanics exactly describes complex systems
% like the observers and their measuring devices, one needs to either give up
% the view that there is a single reality, or the expectation of consistency between observations made by different agents. 
% }
% A
% single-world interpretation is any interpretation that asserts, for a
 %quantum measurement with multiple possible outcomes, that just one
 %outcome and its corresponding reduction of the state actually occurs.

 More precisely, they proved that the three following conditions are incompatible:  
% \begin{itemize}
% \item{Q}  -- the theory is described by standard quantum mechanics. This includes the use of the Born rule to say that certain outcomes occur with probability one. This condition also includes an ideal, unitary, quantum theory (if only an implicit assumption in their work~\cite{delRio2018arXiv}),
% \item{C} -- the theory is self consistent. By this, one assumes that the outcomes observed by independent agents agree with each other,
% \item{S}  -- the theory corresponds to a single-world description, where each agent only observes a single outcome.
% \end{itemize}
%\newline
 (Q)  -- the theory is described by standard quantum mechanics. This includes the use of the Born rule to predict that certain outcomes occur with probability one. This condition also presumes ideal, unitary, quantum theory (if only an implicit assumption in their work~\cite{delRio2018arXiv}); %\newline
 (C)  -- the theory is self-consistent. By this, one assumes that predictions done by independent agents agree with each other; %\newline
 (S)  -- the theory corresponds to a single-world description, where each agent only observes a single outcome in a measurement.
  
 To prove the theorem, they consider two observers Alice and Bob that
 perform measurements in their labs $S_A$ and $S_B$. The measurements of Bob
 depend on the outcomes observed by Alice. The measured quantum
 systems, as well as Alice and Bob, their measuring instruments and
 all the systems in their laboratories that become entangled with the
 measuring instruments in registering and recording the outcomes of
 the quantum measurements, including the
 entangled environments, are just two big many-body quantum systems $S_A$
 and $S_B$, which are assumed to be completely isolated from each other
 after Bob receives Alice’s information.  Consider two super-observers,
 Wigner and Friend, with vast technological abilities, who 
 measure a super-observable X of $S_A$ and a super-observable Y of $S_B$.

 As the evolution of $S_A$ and $S_B$ according to ordinary quantum
 mechanics is unitary (Q), the authors are able to show 
 that, for a particular
 situation, there is no consistent story that includes observers
 and super-observers: a pair of outcomes with finite probability,
 according to quantum mechanics, of the super-observers’ measurements
 on the composite observer system is inconsistent with the observers
 obtaining definite (single) outcomes for their measurements.
 Note that their no-go theorem does not give a hint as to which of the three conditions, or extra implicit assumption~\cite{FrauchigerRenner2016,
 delRio2018arXiv,BlogAaronson}, needs to be dropped.

 Following Bub's notation~\cite{Bub2018} the essence of the argument can
 be summarized as follows: while for Alice and Bob the final state
 after their measurements is given by a density matrix,
 \begin{eqnarray}
   \label{eq:rhoeventFR}
   \rho &=& \frac{1}{3} \vert 0\rangle_B \vert h\rangle_A\langle
   h\vert_A \langle 0\vert_B + \frac{1}{3} \vert 0\rangle_B \vert
            t\rangle_A \langle t\vert_A \langle 0\vert_B\nonumber\\
        && +\frac{1}{3}
   \vert 1\rangle_B \vert t\rangle_A \langle t\vert_A \langle 1\vert_B,
 \end{eqnarray}
the super-observers, due to the unitary evolution, will assign
 a pure state for the complete system
 \begin{equation}
   \label{eq:2}
   \vert \psi\rangle = \frac{1}{\sqrt{3}} \left(
     \vert h\rangle_A \vert 0\rangle_B + \vert t\rangle_A \vert
     0\rangle_B+
     \vert t\rangle_A \vert 1 \rangle _B\right).
 \end{equation}
Frauchiger and Renner prove that following the storyline of each of the agents involved leads to contradictions between their observations.

However, if we take into account the loss of unitarity due to 
time and length uncertainties,
the final state for  Alice, Bob and the super-observers
will coincide and take the form~\eqref{eq:rhoeventFR}. Therefore, subsequent
measurements will not lead to any contradiction. 
As a matter of fact, since our definition of event rests on the fact that there are instances in which the state becomes physically identical to a classical probability distribution, as in~\eqref{eq:rhoeventFR}, the picture put forward is thus as consistent as classical probability theory.

 The approach that we have presented here may be also viewed as the way of making quantum mechanics compatible with a consistent
 single-world interpretation by including the loss of unitarity
 induced by fundamental limitations on the measurement of time and length.
 %the use of realistic clocks.
The inclusion of such limitations due to quantum and gravitational effects allows showing that, if one incorporates in quantum theory these effects and consider the corresponding theory (QG), then this extended theory satisfies the conditions (S) and (C) as well. 
%that generalize the F-R definitions.(REF). 

%Note that the theorem of Frauchiger and Renner still holds for the underlying unitary theory in ideal $t$. In such frame, where (Q) holds, it implies that one cannot have (C) and (S) as well. 
%However, limitations from gravitational and quantum mechanical source would make this theory physically inaccessible. In the physically accessible realm events take place, and (QG), (S) and (C) are compatible.  
%Even though the evolution in the ideal time t is unitary the FR  theorem do not apply in  an ideal realm. 
%In fact, the time in which an event occurs is a relational notion that depends on  events in physical clocks. As we have shown it does not make sense to say that an event occurs at a given value of the ideal time t and therefore C and S that refer to events do not make sense in an ideal realm

\section{Discussion}
\label{sec:discussion}

Quantum theory involves a huge leap from the understanding of
preceding theories, giving a central role to the act of measurement.
This holds irrespective of the philosophical standpoint that one
takes: physical predictions following a measurement are different than
before the measurement
Our view is that the understanding of quantum theory is not complete
until a proper characterization of the process of measurement is at hand. 
One that unequivocally
defines the process by which a quantum system undergoes the change
\begin{align}
\rho_\sys \Longrightarrow \rho_\sys^\ev,
\end{align} 
that is, from a coherent superposition to a state that can be
attributed definite outcomes.  Environmental decoherence gets one very close to the
above, identifying instances in which the transition occurs \emph{for
  all practical purposes}. However, in our opinion these sort of
solutions are not enough, since at best one ends up with a subjective,
and fuzzy, notion of event.  There are protocols that serve to flesh
out the fact that the `decoherence solution' is apparent. In
particular, in this paper we consider a set of global protocols for
which the predictions for $\rho_\sys$ and $\rho_\sys^\ev$ differ
within unitary quantum mechanics.

Naturally, we are not alone in this criticism, and endless work has
been devoted to different modifications and/or re-interpretations of
the theory to attempt such understanding.  Many of these attempts
involve trade-offs, for example giving up the possibility of a
single-world description in exchange for keeping the formalism of
quantum theory intact~\cite{everett1957relative, dewitt1970quantum,
  saunders2010many,Lev-manyworlds-encyclopedia}, reinterpreting the
existing theory by willing to lose the notion of
objectivity~\cite{FuchsRevModPhys2013,FuchsAmJPh2014}, accepting
ad-hoc modifications to quantum theory in order to restore a
single-world picture~\cite{Diosi1984,GRW1986, diosi1988continuous,
  PearlePRA1989,GhirardiPRA1990, gisin1989stochastic,
  penrose1996gravity, Gisinarxiv2017}, or even a combination of all of
the above~\cite{MuellerarXiv2017}.  While we see benefits to all of
these approaches, here we attempt a construction of a realist
description, with well defined notion of events, and with minimal,
physically grounded, premises.

Our starting point is to assume that the laws of the universe do not allow for arbitrarily precise measurements. 
The argument behind this is simple. When one considers physical mechanisms that allow to measure a physical quantity, say time, one finds that quantum mechanics leads to uncertainties in such measurements. 
Typically, these uncertainties can be made small by designing systems that suffer less from quantum mechanical effects, for example by increasing energy and/or size. 
However, this procedure clashes with general relativity, which dictates limits on the energetic content within a region. 
Assuming that both of these general traits will survive in a fundamental theory that combines quantum mechanics and general relativity, one invariably ends up with fundamental uncertainties. 
These minuscule errors on the measurement of time and length intervals can have far reaching consequences on what one can interpret quantum theory to be about.

Fundamental uncertainties in time lead to a loss of coherence in the energy basis. In this paper we proved that this loss of coherence
is enough to rule out a large set of global protocols that allow to verify that no objective event occurs within unitary quantum theory.
In our approach, these instances, in which the predictions become indistinguishable to the case in which the system is in a classically interpretable state, are the events. 
As such, we paint a picture with a clear cut definition of when events take place and when they do not, on a pointer basis determined by the interplay of the different Hamiltonians involved in the problem, and with a well defined timescale, for which we gave estimates in a paradigmatic model of a particle delocalized in space. Note, in particular, that our notion of event needs no reference to observers. The condition for the event to happen or not is uniquely defined by the state of the system, and the limits that nature imposes on measurements.
The events defined in this way completely characterize the physical
quantities of the system, and possibly the environment, that take definite values~\cite{GGPPaxiomatic2011}. 
%XXXXX
%The event is not only characterized by something that occurs on the system of interest, but also on the environment. In pa
%XXXXX

Summarizing, the previous analysis suggests the following ontology: the classical world is composed by of events. 
An event takes place when the state of a closed system, -i.e a system for  which its state gives a complete description- takes the form of a statistical mixture. Most states that quantum systems can be found in, in particular coherent superpositions over sets of states, only describe potentialities to produce events. 
On the other hand, statistical mixtures of closed systems are classically interpretable states. 
These states can be thought of as a classical statistical mixture of states corresponding to different outcomes. 
Quantum mechanics provides the probabilities for these, that occur randomly. It has no information about which of the possible outcome has occurred. 
When this condition is satisfied we know that an event occurred but we ignore which one it was. 
The collapse of the wave function when an event is observed is nothing but the actualization of the information that we posses about events that have already occurred.

At this point one would naturally ask: ``isn't the \emph{and/or} problem still present?''. 
That is, have we adequately explained the transition from a superposition to a single outcome?
We believe this to no longer be a problem once \emph{all physical predictions are as is the system is described by a classical probability distribution}. 
If from the beginnings of quantum theory the founding fathers had found that there are instances in which a quantum system is describable by classical probability distributions, then the and/or question would have never arisen, the same way it doesn't come up when tossing a coin.

Note that this is not simply a philosophical re-interpretation of quantum theory. Our predictions are testable, and the Montevideo Interpretation is falsifiable. 
Should experiments searching for deviations from unitary evolution
rule out the fundamental loss of coherence introduced in
Sec.~\ref{sec:fundamentaluncertainties}, or ever more precise clocks
were conceived, then our approach would be proven wrong~\cite{FrowisRevModPhys2018}. 
Modifications of experiments such as~\cite{BassiPRL2017} to probe decoherence in energy would thus be extremely interesting.

%\subsection{Acknowledgments}

\vspace{12pt}

\emph{Acknowledgments ---} 
This work was supported in part by Grant
No. NSF-PHY-1603630, funds of the Hearne Institute for Theoretical
Physics, CCT-LSU, and Pedeciba and Fondo Clemente Estable
FCE\_1\_2014\_1\_103803, and the John Templeton Foundation.

\bibliography{references}

\begin{thebibliography}{74}%
\makeatletter
\providecommand \@ifxundefined [1]{%
 \@ifx{#1\undefined}
}%
\providecommand \@ifnum [1]{%
 \ifnum #1\expandafter \@firstoftwo
 \else \expandafter \@secondoftwo
 \fi
}%
\providecommand \@ifx [1]{%
 \ifx #1\expandafter \@firstoftwo
 \else \expandafter \@secondoftwo
 \fi
}%
\providecommand \natexlab [1]{#1}%
\providecommand \enquote  [1]{``#1''}%
\providecommand \bibnamefont  [1]{#1}%
\providecommand \bibfnamefont [1]{#1}%
\providecommand \citenamefont [1]{#1}%
\providecommand \href@noop [0]{\@secondoftwo}%
\providecommand \href [0]{\begingroup \@sanitize@url \@href}%
\providecommand \@href[1]{\@@startlink{#1}\@@href}%
\providecommand \@@href[1]{\endgroup#1\@@endlink}%
\providecommand \@sanitize@url [0]{\catcode `\\12\catcode `\$12\catcode
  `\&12\catcode `\#12\catcode `\^12\catcode `\_12\catcode `\%12\relax}%
\providecommand \@@startlink[1]{}%
\providecommand \@@endlink[0]{}%
\providecommand \url  [0]{\begingroup\@sanitize@url \@url }%
\providecommand \@url [1]{\endgroup\@href {#1}{\urlprefix }}%
\providecommand \urlprefix  [0]{URL }%
\providecommand \Eprint [0]{\href }%
\providecommand \doibase [0]{http://dx.doi.org/}%
\providecommand \selectlanguage [0]{\@gobble}%
\providecommand \bibinfo  [0]{\@secondoftwo}%
\providecommand \bibfield  [0]{\@secondoftwo}%
\providecommand \translation [1]{[#1]}%
\providecommand \BibitemOpen [0]{}%
\providecommand \bibitemStop [0]{}%
\providecommand \bibitemNoStop [0]{.\EOS\space}%
\providecommand \EOS [0]{\spacefactor3000\relax}%
\providecommand \BibitemShut  [1]{\csname bibitem#1\endcsname}%
\let\auto@bib@innerbib\@empty
%</preamble>
\bibitem [{\citenamefont {{Spekkens}}(2007)}]{spekkens2007}%
  \BibitemOpen
  \bibfield  {author} {\bibinfo {author} {\bibfnamefont {R.~W.}\ \bibnamefont
  {{Spekkens}}},\ }\bibfield  {title} {\enquote {\bibinfo {title} {{Evidence
  for the epistemic view of quantum states: A toy theory}},}\ }\href {\doibase
  10.1103/PhysRevA.75.032110} {\bibfield  {journal} {\bibinfo  {journal}
  {\pra}\ }\textbf {\bibinfo {volume} {75}},\ \bibinfo {eid} {032110} (\bibinfo
  {year} {2007})},\ \Eprint {http://arxiv.org/abs/quant-ph/0401052}
  {quant-ph/0401052} \BibitemShut {NoStop}%
\bibitem [{\citenamefont {Leifer}(2014)}]{Leifer2014}%
  \BibitemOpen
  \bibfield  {author} {\bibinfo {author} {\bibfnamefont {M.}~\bibnamefont
  {Leifer}},\ }\bibfield  {title} {\enquote {\bibinfo {title} {Is the quantum
  state real? an extended review of $\psi$-ontology theorems},}\ }\href@noop {}
  {\bibfield  {journal} {\bibinfo  {journal} {Quanta}\ }\textbf {\bibinfo
  {volume} {3}},\ \bibinfo {pages} {67--155} (\bibinfo {year}
  {2014})}\BibitemShut {NoStop}%
\bibitem [{\citenamefont {{Gambini}}\ \emph {et~al.}(2010)\citenamefont
  {{Gambini}}, \citenamefont {{Pintos}},\ and\ \citenamefont
  {{Pullin}}}]{GGPPundecidability2010}%
  \BibitemOpen
  \bibfield  {author} {\bibinfo {author} {\bibfnamefont {R.}~\bibnamefont
  {{Gambini}}}, \bibinfo {author} {\bibfnamefont {L.~P.~G.}\ \bibnamefont
  {{Pintos}}}, \ and\ \bibinfo {author} {\bibfnamefont {J.}~\bibnamefont
  {{Pullin}}},\ }\bibfield  {title} {\enquote {\bibinfo {title}
  {{Undecidability and the Problem of Outcomes in Quantum Measurements}},}\
  }\href {\doibase 10.1007/s10701-009-9376-8} {\bibfield  {journal} {\bibinfo
  {journal} {Foundations of Physics}\ }\textbf {\bibinfo {volume} {40}},\
  \bibinfo {pages} {93--115} (\bibinfo {year} {2010})},\ \Eprint
  {http://arxiv.org/abs/0905.4222} {arXiv:0905.4222 [quant-ph]} \BibitemShut
  {NoStop}%
\bibitem [{\citenamefont {{Gambini}}\ \emph
  {et~al.}(2011{\natexlab{a}})\citenamefont {{Gambini}}, \citenamefont
  {{Garc{\'{\i}}a-Pintos}},\ and\ \citenamefont
  {{Pullin}}}]{GGPPcriterionevents2011}%
  \BibitemOpen
  \bibfield  {author} {\bibinfo {author} {\bibfnamefont {R.}~\bibnamefont
  {{Gambini}}}, \bibinfo {author} {\bibfnamefont {L.~P.}\ \bibnamefont
  {{Garc{\'{\i}}a-Pintos}}}, \ and\ \bibinfo {author} {\bibfnamefont
  {J.}~\bibnamefont {{Pullin}}},\ }\bibfield  {title} {\enquote {\bibinfo
  {title} {{Undecidability as Solution to the Problem of Measurement:.
  Fundamental Criterion for the Production of Events}},}\ }\href {\doibase
  10.1142/S0218271811019104} {\bibfield  {journal} {\bibinfo  {journal}
  {International Journal of Modern Physics D}\ }\textbf {\bibinfo {volume}
  {20}},\ \bibinfo {pages} {909--918} (\bibinfo {year} {2011}{\natexlab{a}})},\
  \Eprint {http://arxiv.org/abs/1009.3817} {arXiv:1009.3817 [quant-ph]}
  \BibitemShut {NoStop}%
\bibitem [{\citenamefont {Gambini}\ and\ \citenamefont
  {Pullin}(2009)}]{GPfaq2009}%
  \BibitemOpen
  \bibfield  {author} {\bibinfo {author} {\bibfnamefont {R.}~\bibnamefont
  {Gambini}}\ and\ \bibinfo {author} {\bibfnamefont {J.}~\bibnamefont
  {Pullin}},\ }\bibfield  {title} {\enquote {\bibinfo {title} {The montevideo
  interpretation of quantum mechanics: frequently asked questions},}\ }\href
  {http://stacks.iop.org/1742-6596/174/i=1/a=012003} {\bibfield  {journal}
  {\bibinfo  {journal} {Journal of Physics: Conference Series}\ }\textbf
  {\bibinfo {volume} {174}},\ \bibinfo {pages} {012003} (\bibinfo {year}
  {2009})}\BibitemShut {NoStop}%
\bibitem [{\citenamefont {{Gambini}}\ \emph
  {et~al.}(2011{\natexlab{b}})\citenamefont {{Gambini}}, \citenamefont
  {{Garc{\'{\i}}a-Pintos}},\ and\ \citenamefont
  {{Pullin}}}]{GGPPaxiomatic2011}%
  \BibitemOpen
  \bibfield  {author} {\bibinfo {author} {\bibfnamefont {R.}~\bibnamefont
  {{Gambini}}}, \bibinfo {author} {\bibfnamefont {L.~P.}\ \bibnamefont
  {{Garc{\'{\i}}a-Pintos}}}, \ and\ \bibinfo {author} {\bibfnamefont
  {J.}~\bibnamefont {{Pullin}}},\ }\bibfield  {title} {\enquote {\bibinfo
  {title} {{An axiomatic formulation of the Montevideo interpretation of
  quantum mechanics}},}\ }\href {\doibase 10.1016/j.shpsb.2011.10.002}
  {\bibfield  {journal} {\bibinfo  {journal} {Studies in the History and
  Philosophy of Modern Physics}\ }\textbf {\bibinfo {volume} {42}},\ \bibinfo
  {pages} {256--263} (\bibinfo {year} {2011}{\natexlab{b}})},\ \Eprint
  {http://arxiv.org/abs/1002.4209} {arXiv:1002.4209 [quant-ph]} \BibitemShut
  {NoStop}%
\bibitem [{\citenamefont {Gambini}\ and\ \citenamefont
  {Pullin}(2018)}]{gambini2015montevideo}%
  \BibitemOpen
  \bibfield  {author} {\bibinfo {author} {\bibfnamefont {R.}~\bibnamefont
  {Gambini}}\ and\ \bibinfo {author} {\bibfnamefont {J.}~\bibnamefont
  {Pullin}},\ }\bibfield  {title} {\enquote {\bibinfo {title} {The montevideo
  interpretation of quantum mechanics: A short review},}\ }\href@noop {}
  {\bibfield  {journal} {\bibinfo  {journal} {Entropy}\ }\textbf {\bibinfo
  {volume} {20}} (\bibinfo {year} {2018})}\BibitemShut {NoStop}%
\bibitem [{\citenamefont {Zeh}(1970)}]{Zeh1970}%
  \BibitemOpen
  \bibfield  {author} {\bibinfo {author} {\bibfnamefont {H.~D.}\ \bibnamefont
  {Zeh}},\ }\bibfield  {title} {\enquote {\bibinfo {title} {On the
  interpretation of measurement in quantum theory},}\ }\href {\doibase
  10.1007/BF00708656} {\bibfield  {journal} {\bibinfo  {journal} {Foundations
  of Physics}\ }\textbf {\bibinfo {volume} {1}},\ \bibinfo {pages} {69--76}
  (\bibinfo {year} {1970})}\BibitemShut {NoStop}%
\bibitem [{\citenamefont {Paz}\ and\ \citenamefont
  {Zurek}(1999)}]{PazpointerbasisPRL1999}%
  \BibitemOpen
  \bibfield  {author} {\bibinfo {author} {\bibfnamefont {J.~P.}\ \bibnamefont
  {Paz}}\ and\ \bibinfo {author} {\bibfnamefont {W.~H.}\ \bibnamefont
  {Zurek}},\ }\bibfield  {title} {\enquote {\bibinfo {title} {Quantum limit of
  decoherence: Environment induced superselection of energy eigenstates},}\
  }\href {\doibase 10.1103/PhysRevLett.82.5181} {\bibfield  {journal} {\bibinfo
   {journal} {Phys. Rev. Lett.}\ }\textbf {\bibinfo {volume} {82}},\ \bibinfo
  {pages} {5181--5185} (\bibinfo {year} {1999})}\BibitemShut {NoStop}%
\bibitem [{\citenamefont {Schlosshauer}(2005)}]{schlosshauer2005review}%
  \BibitemOpen
  \bibfield  {author} {\bibinfo {author} {\bibfnamefont {M.}~\bibnamefont
  {Schlosshauer}},\ }\bibfield  {title} {\enquote {\bibinfo {title}
  {Decoherence, the measurement problem, and interpretations of quantum
  mechanics},}\ }\href {\doibase 10.1103/RevModPhys.76.1267} {\bibfield
  {journal} {\bibinfo  {journal} {Rev. Mod. Phys.}\ }\textbf {\bibinfo {volume}
  {76}},\ \bibinfo {pages} {1267--1305} (\bibinfo {year} {2005})}\BibitemShut
  {NoStop}%
\bibitem [{\citenamefont {{Schlosshauer}}(2014)}]{schlosshauer2014review}%
  \BibitemOpen
  \bibfield  {author} {\bibinfo {author} {\bibfnamefont {M.}~\bibnamefont
  {{Schlosshauer}}},\ }\bibfield  {title} {\enquote {\bibinfo {title} {{The
  quantum-to-classical transition and decoherence}},}\ }\href@noop {}
  {\bibfield  {journal} {\bibinfo  {journal} {ArXiv e-prints}\ } (\bibinfo
  {year} {2014})},\ \Eprint {http://arxiv.org/abs/1404.2635} {arXiv:1404.2635
  [quant-ph]} \BibitemShut {NoStop}%
\bibitem [{\citenamefont {Zurek}(1991)}]{zurek1991}%
  \BibitemOpen
  \bibfield  {author} {\bibinfo {author} {\bibfnamefont {W.~H.}\ \bibnamefont
  {Zurek}},\ }\bibfield  {title} {\enquote {\bibinfo {title} {Decoherence and
  the transition from quantum to classical},}\ }\href@noop {} {\bibfield
  {journal} {\bibinfo  {journal} {Physics today}\ }\textbf {\bibinfo {volume}
  {44}},\ \bibinfo {pages} {36--44} (\bibinfo {year} {1991})}\BibitemShut
  {NoStop}%
\bibitem [{\citenamefont {Zurek}(2003)}]{Zurek2003}%
  \BibitemOpen
  \bibfield  {author} {\bibinfo {author} {\bibfnamefont {W.~H.}\ \bibnamefont
  {Zurek}},\ }\bibfield  {title} {\enquote {\bibinfo {title} {Decoherence,
  einselection, and the quantum origins of the classical},}\ }\href {\doibase
  10.1103/RevModPhys.75.715} {\bibfield  {journal} {\bibinfo  {journal} {Rev.
  Mod. Phys.}\ }\textbf {\bibinfo {volume} {75}},\ \bibinfo {pages} {715--775}
  (\bibinfo {year} {2003})}\BibitemShut {NoStop}%
\bibitem [{\citenamefont {Riedel}\ and\ \citenamefont
  {Zurek}(2010)}]{riedel2010}%
  \BibitemOpen
  \bibfield  {author} {\bibinfo {author} {\bibfnamefont {C.~J.}\ \bibnamefont
  {Riedel}}\ and\ \bibinfo {author} {\bibfnamefont {W.~H.}\ \bibnamefont
  {Zurek}},\ }\bibfield  {title} {\enquote {\bibinfo {title} {Quantum darwinism
  in an everyday environment: Huge redundancy in scattered photons},}\
  }\href@noop {} {\bibfield  {journal} {\bibinfo  {journal} {Physical review
  letters}\ }\textbf {\bibinfo {volume} {105}},\ \bibinfo {pages} {020404}
  (\bibinfo {year} {2010})}\BibitemShut {NoStop}%
\bibitem [{\citenamefont {Schlosshauer}(2007)}]{schlosshauer2007book}%
  \BibitemOpen
  \bibfield  {author} {\bibinfo {author} {\bibfnamefont {M.}~\bibnamefont
  {Schlosshauer}},\ }\href@noop {} {\emph {\bibinfo {title} {Decoherence: and
  the quantum-to-classical transition}}}\ (\bibinfo  {publisher} {Springer
  Science \& Business Media},\ \bibinfo {year} {2007})\BibitemShut {NoStop}%
\bibitem [{\citenamefont {Joos}\ \emph {et~al.}(2013)\citenamefont {Joos},
  \citenamefont {Zeh}, \citenamefont {Kiefer}, \citenamefont {Giulini},
  \citenamefont {Kupsch},\ and\ \citenamefont {Stamatescu}}]{joos2013}%
  \BibitemOpen
  \bibfield  {author} {\bibinfo {author} {\bibfnamefont {E.}~\bibnamefont
  {Joos}}, \bibinfo {author} {\bibfnamefont {H.~D.}\ \bibnamefont {Zeh}},
  \bibinfo {author} {\bibfnamefont {C.}~\bibnamefont {Kiefer}}, \bibinfo
  {author} {\bibfnamefont {D.~J.~W.}\ \bibnamefont {Giulini}}, \bibinfo
  {author} {\bibfnamefont {J.}~\bibnamefont {Kupsch}}, \ and\ \bibinfo {author}
  {\bibfnamefont {I.}~\bibnamefont {Stamatescu}},\ }\href@noop {} {\emph
  {\bibinfo {title} {Decoherence and the appearance of a classical world in
  quantum theory}}}\ (\bibinfo  {publisher} {Springer Science \& Business
  Media},\ \bibinfo {year} {2013})\BibitemShut {NoStop}%
\bibitem [{\citenamefont {Wigner}(1967)}]{BookWigner1967}%
  \BibitemOpen
  \bibfield  {author} {\bibinfo {author} {\bibfnamefont {E.~P.}\ \bibnamefont
  {Wigner}},\ }\href@noop {} {\emph {\bibinfo {title} {Symmetries and
  reflections: Scientific essays of Eugene P. Wigner}}}\ (\bibinfo  {publisher}
  {Indiana University Press Bloomington},\ \bibinfo {year} {1967})\BibitemShut
  {NoStop}%
\bibitem [{\citenamefont {d'Espagnat}(1995)}]{dEspagnat1995}%
  \BibitemOpen
  \bibfield  {author} {\bibinfo {author} {\bibfnamefont {B.}~\bibnamefont
  {d'Espagnat}},\ }\bibfield  {title} {\enquote {\bibinfo {title} {Veiled
  reality an analysis of present-day quantum mechanical concepts},}\
  }\href@noop {} {\  (\bibinfo {year} {1995})}\BibitemShut {NoStop}%
\bibitem [{\citenamefont {Frauchiger}\ and\ \citenamefont
  {Renner}(2018)}]{FrauchigerRenner2016}%
  \BibitemOpen
  \bibfield  {author} {\bibinfo {author} {\bibfnamefont {D.}~\bibnamefont
  {Frauchiger}}\ and\ \bibinfo {author} {\bibfnamefont {R.}~\bibnamefont
  {Renner}},\ }\bibfield  {title} {\enquote {\bibinfo {title} {{Quantum theory
  cannot consistently describe the use of itself}},}\ }\href {\doibase
  10.1038/s41467-018-05739-8} {\bibfield  {journal} {\bibinfo  {journal}
  {Nature Communications}\ }\textbf {\bibinfo {volume} {9}},\ \bibinfo {pages}
  {3711} (\bibinfo {year} {2018})}\BibitemShut {NoStop}%
\bibitem [{\citenamefont {Jalabert}\ and\ \citenamefont
  {Pastawski}(2001)}]{JalabertPastawskiPRL2001}%
  \BibitemOpen
  \bibfield  {author} {\bibinfo {author} {\bibfnamefont {R.~A.}\ \bibnamefont
  {Jalabert}}\ and\ \bibinfo {author} {\bibfnamefont {H.~M.}\ \bibnamefont
  {Pastawski}},\ }\bibfield  {title} {\enquote {\bibinfo {title}
  {Environment-independent decoherence rate in classically chaotic systems},}\
  }\href {\doibase 10.1103/PhysRevLett.86.2490} {\bibfield  {journal} {\bibinfo
   {journal} {Phys. Rev. Lett.}\ }\textbf {\bibinfo {volume} {86}},\ \bibinfo
  {pages} {2490--2493} (\bibinfo {year} {2001})}\BibitemShut {NoStop}%
\bibitem [{\citenamefont {Petitjean}\ and\ \citenamefont
  {Jacquod}(2006)}]{PetitjeanPRL2006}%
  \BibitemOpen
  \bibfield  {author} {\bibinfo {author} {\bibfnamefont {C.}~\bibnamefont
  {Petitjean}}\ and\ \bibinfo {author} {\bibfnamefont {Ph.}\ \bibnamefont
  {Jacquod}},\ }\bibfield  {title} {\enquote {\bibinfo {title} {Quantum
  reversibility and echoes in interacting systems},}\ }\href {\doibase
  10.1103/PhysRevLett.97.124103} {\bibfield  {journal} {\bibinfo  {journal}
  {Phys. Rev. Lett.}\ }\textbf {\bibinfo {volume} {97}},\ \bibinfo {pages}
  {124103} (\bibinfo {year} {2006})}\BibitemShut {NoStop}%
\bibitem [{\citenamefont {Goussev}\ \emph {et~al.}(2012)\citenamefont
  {Goussev}, \citenamefont {Jalabert}, \citenamefont {Pastawski},\ and\
  \citenamefont {Wisniacki}}]{LoschmidtEcho2012}%
  \BibitemOpen
  \bibfield  {author} {\bibinfo {author} {\bibfnamefont {A.}~\bibnamefont
  {Goussev}}, \bibinfo {author} {\bibfnamefont {R.~A.}\ \bibnamefont
  {Jalabert}}, \bibinfo {author} {\bibfnamefont {H.~M.}\ \bibnamefont
  {Pastawski}}, \ and\ \bibinfo {author} {\bibfnamefont {D.~Ariel}\
  \bibnamefont {Wisniacki}},\ }\bibfield  {title} {\enquote {\bibinfo {title}
  {{L}oschmidt echo},}\ }\href {\doibase 10.4249/scholarpedia.11687} {\bibfield
   {journal} {\bibinfo  {journal} {Scholarpedia}\ }\textbf {\bibinfo {volume}
  {7}},\ \bibinfo {pages} {11687} (\bibinfo {year} {2012})},\ \bibinfo {note}
  {revision \#127578}\BibitemShut {NoStop}%
\bibitem [{\citenamefont {Peres}(1984)}]{Peres84}%
  \BibitemOpen
  \bibfield  {author} {\bibinfo {author} {\bibfnamefont {A.}~\bibnamefont
  {Peres}},\ }\bibfield  {title} {\enquote {\bibinfo {title} {Stability of
  quantum motion in chaotic and regular systems},}\ }\href {\doibase
  10.1103/PhysRevA.30.1610} {\bibfield  {journal} {\bibinfo  {journal} {Phys.
  Rev. A}\ }\textbf {\bibinfo {volume} {30}},\ \bibinfo {pages} {1610--1615}
  (\bibinfo {year} {1984})}\BibitemShut {NoStop}%
\bibitem [{\citenamefont {Navascu\'es}(2018)}]{NavascuesPRX2018}%
  \BibitemOpen
  \bibfield  {author} {\bibinfo {author} {\bibfnamefont {M.}~\bibnamefont
  {Navascu\'es}},\ }\bibfield  {title} {\enquote {\bibinfo {title} {Resetting
  uncontrolled quantum systems},}\ }\href {\doibase 10.1103/PhysRevX.8.031008}
  {\bibfield  {journal} {\bibinfo  {journal} {Phys. Rev. X}\ }\textbf {\bibinfo
  {volume} {8}},\ \bibinfo {pages} {031008} (\bibinfo {year}
  {2018})}\BibitemShut {NoStop}%
\bibitem [{\citenamefont {Saunders}\ \emph {et~al.}(2010)\citenamefont
  {Saunders}, \citenamefont {Barrett}, \citenamefont {Kent},\ and\
  \citenamefont {Wallace}}]{saunders2010many}%
  \BibitemOpen
  \bibfield  {author} {\bibinfo {author} {\bibfnamefont {S.}~\bibnamefont
  {Saunders}}, \bibinfo {author} {\bibfnamefont {J.}~\bibnamefont {Barrett}},
  \bibinfo {author} {\bibfnamefont {A.}~\bibnamefont {Kent}}, \ and\ \bibinfo
  {author} {\bibfnamefont {D.}~\bibnamefont {Wallace}},\ }\href@noop {} {\emph
  {\bibinfo {title} {Many worlds?: Everett, quantum theory, \& reality}}}\
  (\bibinfo  {publisher} {Oxford University Press},\ \bibinfo {year}
  {2010})\BibitemShut {NoStop}%
\bibitem [{\citenamefont {Salecker}\ and\ \citenamefont
  {Wigner}(1958)}]{SaleckerWignerPhysRev1958}%
  \BibitemOpen
  \bibfield  {author} {\bibinfo {author} {\bibfnamefont {H.}~\bibnamefont
  {Salecker}}\ and\ \bibinfo {author} {\bibfnamefont {E.~P.}\ \bibnamefont
  {Wigner}},\ }\bibfield  {title} {\enquote {\bibinfo {title} {Quantum
  limitations of the measurement of space-time distances},}\ }\href {\doibase
  10.1103/PhysRev.109.571} {\bibfield  {journal} {\bibinfo  {journal} {Phys.
  Rev.}\ }\textbf {\bibinfo {volume} {109}},\ \bibinfo {pages} {571--577}
  (\bibinfo {year} {1958})}\BibitemShut {NoStop}%
\bibitem [{\citenamefont {Peres}(1980)}]{PeresAmJPhys80}%
  \BibitemOpen
  \bibfield  {author} {\bibinfo {author} {\bibfnamefont {A.}~\bibnamefont
  {Peres}},\ }\bibfield  {title} {\enquote {\bibinfo {title} {Measurement of
  time by quantum clocks},}\ }\href@noop {} {\bibfield  {journal} {\bibinfo
  {journal} {American Journal of Physics}\ }\textbf {\bibinfo {volume} {48}},\
  \bibinfo {pages} {552--557} (\bibinfo {year} {1980})}\BibitemShut {NoStop}%
\bibitem [{\citenamefont {Peres}(2006)}]{PeresBook2006}%
  \BibitemOpen
  \bibfield  {author} {\bibinfo {author} {\bibfnamefont {A.}~\bibnamefont
  {Peres}},\ }\href@noop {} {\emph {\bibinfo {title} {Quantum theory: concepts
  and methods}}},\ Vol.~\bibinfo {volume} {57}\ (\bibinfo  {publisher}
  {Springer Science \& Business Media},\ \bibinfo {year} {2006})\BibitemShut
  {NoStop}%
\bibitem [{\citenamefont {Aharonov}\ \emph {et~al.}(1998)\citenamefont
  {Aharonov}, \citenamefont {Oppenheim}, \citenamefont {Popescu}, \citenamefont
  {Reznik},\ and\ \citenamefont {Unruh}}]{AharonovPRA98}%
  \BibitemOpen
  \bibfield  {author} {\bibinfo {author} {\bibfnamefont {Y.}~\bibnamefont
  {Aharonov}}, \bibinfo {author} {\bibfnamefont {J.}~\bibnamefont {Oppenheim}},
  \bibinfo {author} {\bibfnamefont {S.}~\bibnamefont {Popescu}}, \bibinfo
  {author} {\bibfnamefont {B.}~\bibnamefont {Reznik}}, \ and\ \bibinfo {author}
  {\bibfnamefont {W.~G.}\ \bibnamefont {Unruh}},\ }\bibfield  {title} {\enquote
  {\bibinfo {title} {Measurement of time of arrival in quantum mechanics},}\
  }\href {\doibase 10.1103/PhysRevA.57.4130} {\bibfield  {journal} {\bibinfo
  {journal} {Phys. Rev. A}\ }\textbf {\bibinfo {volume} {57}},\ \bibinfo
  {pages} {4130--4139} (\bibinfo {year} {1998})}\BibitemShut {NoStop}%
\bibitem [{\citenamefont {Bu\ifmmode~\check{z}\else \v{z}\fi{}ek}\ \emph
  {et~al.}(1999)\citenamefont {Bu\ifmmode~\check{z}\else \v{z}\fi{}ek},
  \citenamefont {Derka},\ and\ \citenamefont {Massar}}]{MassarPRL99}%
  \BibitemOpen
  \bibfield  {author} {\bibinfo {author} {\bibfnamefont {V.}~\bibnamefont
  {Bu\ifmmode~\check{z}\else \v{z}\fi{}ek}}, \bibinfo {author} {\bibfnamefont
  {R.}~\bibnamefont {Derka}}, \ and\ \bibinfo {author} {\bibfnamefont
  {S.}~\bibnamefont {Massar}},\ }\bibfield  {title} {\enquote {\bibinfo {title}
  {Optimal quantum clocks},}\ }\href {\doibase 10.1103/PhysRevLett.82.2207}
  {\bibfield  {journal} {\bibinfo  {journal} {Phys. Rev. Lett.}\ }\textbf
  {\bibinfo {volume} {82}},\ \bibinfo {pages} {2207--2210} (\bibinfo {year}
  {1999})}\BibitemShut {NoStop}%
\bibitem [{\citenamefont {Muga}\ \emph {et~al.}(2007)\citenamefont {Muga},
  \citenamefont {Mayato},\ and\ \citenamefont {Egusquiza}}]{booktimevol1}%
  \BibitemOpen
  \bibfield  {author} {\bibinfo {author} {\bibfnamefont {J.~G.}\ \bibnamefont
  {Muga}}, \bibinfo {author} {\bibfnamefont {R.~Sala}\ \bibnamefont {Mayato}},
  \ and\ \bibinfo {author} {\bibfnamefont {I.~L.}\ \bibnamefont {Egusquiza}},\
  }\href {https://www.springer.com/us/book/9783540734727} {\emph {\bibinfo
  {title} {Time in Quantum Mechanics, Vol. 1}}}\ (\bibinfo  {publisher}
  {Springer, Berlin},\ \bibinfo {year} {2007})\BibitemShut {NoStop}%
\bibitem [{\citenamefont {Muga}\ \emph {et~al.}(2009)\citenamefont {Muga},
  \citenamefont {Ruschhaupt},\ and\ \citenamefont {del Campo}}]{booktimevol2}%
  \BibitemOpen
  \bibfield  {author} {\bibinfo {author} {\bibfnamefont {J.~G.}\ \bibnamefont
  {Muga}}, \bibinfo {author} {\bibfnamefont {A.}~\bibnamefont {Ruschhaupt}}, \
  and\ \bibinfo {author} {\bibfnamefont {A.}~\bibnamefont {del Campo}},\ }\href
  {https://www.springer.com/us/book/9783642031731} {\emph {\bibinfo {title}
  {Time in Quantum Mechanics, Vol. 2}}}\ (\bibinfo  {publisher} {Springer,
  Berlin},\ \bibinfo {year} {2009})\BibitemShut {NoStop}%
\bibitem [{\citenamefont {Khosla}\ and\ \citenamefont
  {Altamirano}(2017)}]{Khosla:2016tss}%
  \BibitemOpen
  \bibfield  {author} {\bibinfo {author} {\bibfnamefont {K.~E.}\ \bibnamefont
  {Khosla}}\ and\ \bibinfo {author} {\bibfnamefont {N.}~\bibnamefont
  {Altamirano}},\ }\bibfield  {title} {\enquote {\bibinfo {title} {Detecting
  gravitational decoherence with clocks: Limits on temporal resolution from a
  classical-channel model of gravity},}\ }\href {\doibase
  10.1103/PhysRevA.95.052116} {\bibfield  {journal} {\bibinfo  {journal} {Phys.
  Rev. A}\ }\textbf {\bibinfo {volume} {95}},\ \bibinfo {pages} {052116}
  (\bibinfo {year} {2017})}\BibitemShut {NoStop}%
\bibitem [{\citenamefont {{Castro Ruiz}}\ \emph {et~al.}(2017)\citenamefont
  {{Castro Ruiz}}, \citenamefont {{Giacomini}},\ and\ \citenamefont
  {{Brukner}}}]{CastroRuizPNAS2017}%
  \BibitemOpen
  \bibfield  {author} {\bibinfo {author} {\bibfnamefont {E.}~\bibnamefont
  {{Castro Ruiz}}}, \bibinfo {author} {\bibfnamefont {F.}~\bibnamefont
  {{Giacomini}}}, \ and\ \bibinfo {author} {\bibfnamefont {{\v
  C}.}~\bibnamefont {{Brukner}}},\ }\bibfield  {title} {\enquote {\bibinfo
  {title} {{Entanglement of quantum clocks through gravity}},}\ }\href
  {\doibase 10.1073/pnas.1616427114} {\bibfield  {journal} {\bibinfo  {journal}
  {Proceedings of the National Academy of Science}\ }\textbf {\bibinfo {volume}
  {114}},\ \bibinfo {pages} {E2303--E2309} (\bibinfo {year} {2017})},\ \Eprint
  {http://arxiv.org/abs/1507.01955} {arXiv:1507.01955 [quant-ph]} \BibitemShut
  {NoStop}%
\bibitem [{\citenamefont {{Vanrietvelde}}\ \emph
  {et~al.}(2018{\natexlab{a}})\citenamefont {{Vanrietvelde}}, \citenamefont
  {{Hoehn}}, \citenamefont {{Giacomini}},\ and\ \citenamefont
  {{Castro-Ruiz}}}]{Hoehn1}%
  \BibitemOpen
  \bibfield  {author} {\bibinfo {author} {\bibfnamefont {A.}~\bibnamefont
  {{Vanrietvelde}}}, \bibinfo {author} {\bibfnamefont {P.~A}\ \bibnamefont
  {{Hoehn}}}, \bibinfo {author} {\bibfnamefont {F.}~\bibnamefont
  {{Giacomini}}}, \ and\ \bibinfo {author} {\bibfnamefont {E.}~\bibnamefont
  {{Castro-Ruiz}}},\ }\bibfield  {title} {\enquote {\bibinfo {title} {{A change
  of perspective: switching quantum reference frames via a perspective-neutral
  framework}},}\ }\href@noop {} {\bibfield  {journal} {\bibinfo  {journal}
  {ArXiv e-prints}\ } (\bibinfo {year} {2018}{\natexlab{a}})},\ \Eprint
  {http://arxiv.org/abs/1809.00556} {arXiv:1809.00556 [quant-ph]} \BibitemShut
  {NoStop}%
\bibitem [{\citenamefont {{Vanrietvelde}}\ \emph
  {et~al.}(2018{\natexlab{b}})\citenamefont {{Vanrietvelde}}, \citenamefont
  {{Hoehn}},\ and\ \citenamefont {{Giacomini}}}]{Hoehn2}%
  \BibitemOpen
  \bibfield  {author} {\bibinfo {author} {\bibfnamefont {A.}~\bibnamefont
  {{Vanrietvelde}}}, \bibinfo {author} {\bibfnamefont {P.~A}\ \bibnamefont
  {{Hoehn}}}, \ and\ \bibinfo {author} {\bibfnamefont {F.}~\bibnamefont
  {{Giacomini}}},\ }\bibfield  {title} {\enquote {\bibinfo {title} {{Switching
  quantum reference frames in the N-body problem and the absence of global
  relational perspectives}},}\ }\href@noop {} {\bibfield  {journal} {\bibinfo
  {journal} {ArXiv e-prints}\ } (\bibinfo {year} {2018}{\natexlab{b}})},\
  \Eprint {http://arxiv.org/abs/1809.05093} {arXiv:1809.05093 [quant-ph]}
  \BibitemShut {NoStop}%
\bibitem [{\citenamefont {{Hoehn}}\ and\ \citenamefont
  {{Vanrietvelde}}(2018)}]{Hoehn3}%
  \BibitemOpen
  \bibfield  {author} {\bibinfo {author} {\bibfnamefont {P.~A}\ \bibnamefont
  {{Hoehn}}}\ and\ \bibinfo {author} {\bibfnamefont {A.}~\bibnamefont
  {{Vanrietvelde}}},\ }\bibfield  {title} {\enquote {\bibinfo {title} {{How to
  switch between relational quantum clocks}},}\ }\href@noop {} {\bibfield
  {journal} {\bibinfo  {journal} {ArXiv e-prints}\ } (\bibinfo {year}
  {2018})},\ \Eprint {http://arxiv.org/abs/1810.04153} {arXiv:1810.04153
  [gr-qc]} \BibitemShut {NoStop}%
\bibitem [{\citenamefont {Malabarba}\ \emph {et~al.}(2015)\citenamefont
  {Malabarba}, \citenamefont {Short},\ and\ \citenamefont
  {Kammerlander}}]{MalabarbaNJPhys2015}%
  \BibitemOpen
  \bibfield  {author} {\bibinfo {author} {\bibfnamefont {A.~S.~L.}\
  \bibnamefont {Malabarba}}, \bibinfo {author} {\bibfnamefont {A.~J.}\
  \bibnamefont {Short}}, \ and\ \bibinfo {author} {\bibfnamefont
  {P.}~\bibnamefont {Kammerlander}},\ }\bibfield  {title} {\enquote {\bibinfo
  {title} {Clock-driven quantum thermal engines},}\ }\href
  {http://stacks.iop.org/1367-2630/17/i=4/a=045027} {\bibfield  {journal}
  {\bibinfo  {journal} {New Journal of Physics}\ }\textbf {\bibinfo {volume}
  {17}},\ \bibinfo {pages} {045027} (\bibinfo {year} {2015})}\BibitemShut
  {NoStop}%
\bibitem [{\citenamefont {{Rankovi{\'c}}}\ \emph {et~al.}(2015)\citenamefont
  {{Rankovi{\'c}}}, \citenamefont {{Liang}},\ and\ \citenamefont
  {{Renner}}}]{RankovicarXiv2015}%
  \BibitemOpen
  \bibfield  {author} {\bibinfo {author} {\bibfnamefont {S.}~\bibnamefont
  {{Rankovi{\'c}}}}, \bibinfo {author} {\bibfnamefont {Y.-C.}\ \bibnamefont
  {{Liang}}}, \ and\ \bibinfo {author} {\bibfnamefont {R.}~\bibnamefont
  {{Renner}}},\ }\bibfield  {title} {\enquote {\bibinfo {title} {{Quantum
  clocks and their synchronisation - the Alternate Ticks Game}},}\ }\href@noop
  {} {\bibfield  {journal} {\bibinfo  {journal} {ArXiv e-prints}\ } (\bibinfo
  {year} {2015})},\ \Eprint {http://arxiv.org/abs/1506.01373} {arXiv:1506.01373
  [quant-ph]} \BibitemShut {NoStop}%
\bibitem [{\citenamefont {{Erker}}\ \emph {et~al.}(2017)\citenamefont
  {{Erker}}, \citenamefont {{Mitchison}}, \citenamefont {{Silva}},
  \citenamefont {{Woods}}, \citenamefont {{Brunner}},\ and\ \citenamefont
  {{Huber}}}]{ErkerPRX2017}%
  \BibitemOpen
  \bibfield  {author} {\bibinfo {author} {\bibfnamefont {P.}~\bibnamefont
  {{Erker}}}, \bibinfo {author} {\bibfnamefont {M.~T.}\ \bibnamefont
  {{Mitchison}}}, \bibinfo {author} {\bibfnamefont {R.}~\bibnamefont
  {{Silva}}}, \bibinfo {author} {\bibfnamefont {M.~P.}\ \bibnamefont
  {{Woods}}}, \bibinfo {author} {\bibfnamefont {N.}~\bibnamefont {{Brunner}}},
  \ and\ \bibinfo {author} {\bibfnamefont {M.}~\bibnamefont {{Huber}}},\
  }\bibfield  {title} {\enquote {\bibinfo {title} {{Autonomous Quantum Clocks:
  Does Thermodynamics Limit Our Ability to Measure Time?}}}\ }\href {\doibase
  10.1103/PhysRevX.7.031022} {\bibfield  {journal} {\bibinfo  {journal}
  {Physical Review X}\ }\textbf {\bibinfo {volume} {7}},\ \bibinfo {eid}
  {031022} (\bibinfo {year} {2017})},\ \Eprint
  {http://arxiv.org/abs/1609.06704} {arXiv:1609.06704 [quant-ph]} \BibitemShut
  {NoStop}%
\bibitem [{\citenamefont {{Woods}}\ \emph {et~al.}(2018)\citenamefont
  {{Woods}}, \citenamefont {{Silva}}, \citenamefont {{P{\"u}tz}}, \citenamefont
  {{Stupar}},\ and\ \citenamefont {{Renner}}}]{WoodsarXiv2018}%
  \BibitemOpen
  \bibfield  {author} {\bibinfo {author} {\bibfnamefont {M.~P.}\ \bibnamefont
  {{Woods}}}, \bibinfo {author} {\bibfnamefont {R.}~\bibnamefont {{Silva}}},
  \bibinfo {author} {\bibfnamefont {G.}~\bibnamefont {{P{\"u}tz}}}, \bibinfo
  {author} {\bibfnamefont {S.}~\bibnamefont {{Stupar}}}, \ and\ \bibinfo
  {author} {\bibfnamefont {R.}~\bibnamefont {{Renner}}},\ }\bibfield  {title}
  {\enquote {\bibinfo {title} {{Quantum clocks are more accurate than classical
  ones}},}\ }\href@noop {} {\bibfield  {journal} {\bibinfo  {journal} {ArXiv
  e-prints}\ } (\bibinfo {year} {2018})},\ \Eprint
  {http://arxiv.org/abs/1806.00491} {arXiv:1806.00491 [quant-ph]} \BibitemShut
  {NoStop}%
\bibitem [{\citenamefont {{Ng}}\ and\ \citenamefont {{van
  Dam}}(1995)}]{NgAnnals1995}%
  \BibitemOpen
  \bibfield  {author} {\bibinfo {author} {\bibfnamefont {Y.~J.}\ \bibnamefont
  {{Ng}}}\ and\ \bibinfo {author} {\bibfnamefont {H.}~\bibnamefont {{van
  Dam}}},\ }\bibfield  {title} {\enquote {\bibinfo {title} {{Limitation to
  Quantum Measurements of Space-Time Distances}},}\ }in\ \href {\doibase
  10.1111/j.1749-6632.1995.tb38998.x} {\emph {\bibinfo {booktitle} {Fundamental
  Problems in Quantum Theory}}},\ \bibinfo {series} {Annals of the New York
  Academy of Sciences}, Vol.\ \bibinfo {volume} {755},\ \bibinfo {editor}
  {edited by\ \bibinfo {editor} {\bibfnamefont {D.~M.}\ \bibnamefont
  {{Greenberger}}}\ and\ \bibinfo {editor} {\bibfnamefont {A.}~\bibnamefont
  {{Zelinger}}}}\ (\bibinfo {year} {1995})\ p.\ \bibinfo {pages} {579},\
  \Eprint {http://arxiv.org/abs/hep-th/9406110} {hep-th/9406110} \BibitemShut
  {NoStop}%
\bibitem [{\citenamefont {Lloyd}(2000)}]{LloydNature2000}%
  \BibitemOpen
  \bibfield  {author} {\bibinfo {author} {\bibfnamefont {S.}~\bibnamefont
  {Lloyd}},\ }\bibfield  {title} {\enquote {\bibinfo {title} {Ultimate physical
  limits to computation},}\ }\href {https://doi.org/10.1038/35023282}
  {\bibfield  {journal} {\bibinfo  {journal} {Nature}\ }\textbf {\bibinfo
  {volume} {406}},\ \bibinfo {pages} {1047} (\bibinfo {year}
  {2000})}\BibitemShut {NoStop}%
\bibitem [{\citenamefont {Baez}\ and\ \citenamefont
  {Olson}(2002)}]{BaezClassquantgrav2002}%
  \BibitemOpen
  \bibfield  {author} {\bibinfo {author} {\bibfnamefont {J.~C.}\ \bibnamefont
  {Baez}}\ and\ \bibinfo {author} {\bibfnamefont {S.~J.}\ \bibnamefont
  {Olson}},\ }\bibfield  {title} {\enquote {\bibinfo {title} {Uncertainty in
  measurements of distance},}\ }\href
  {http://stacks.iop.org/0264-9381/19/i=14/a=101} {\bibfield  {journal}
  {\bibinfo  {journal} {Classical and Quantum Gravity}\ }\textbf {\bibinfo
  {volume} {19}},\ \bibinfo {pages} {L121} (\bibinfo {year}
  {2002})}\BibitemShut {NoStop}%
\bibitem [{\citenamefont {Ng}\ and\ \citenamefont {van
  Dam}(2003)}]{NgClassquantgrav2003}%
  \BibitemOpen
  \bibfield  {author} {\bibinfo {author} {\bibfnamefont {Y.~J.}\ \bibnamefont
  {Ng}}\ and\ \bibinfo {author} {\bibfnamefont {H.}~\bibnamefont {van Dam}},\
  }\bibfield  {title} {\enquote {\bibinfo {title} {Comment on `uncertainty in
  measurements of distance'},}\ }\href
  {http://stacks.iop.org/0264-9381/20/i=2/a=311} {\bibfield  {journal}
  {\bibinfo  {journal} {Classical and Quantum Gravity}\ }\textbf {\bibinfo
  {volume} {20}},\ \bibinfo {pages} {393} (\bibinfo {year} {2003})}\BibitemShut
  {NoStop}%
\bibitem [{\citenamefont {Frenkel}(2010)}]{Frenkel:2010ai}%
  \BibitemOpen
  \bibfield  {author} {\bibinfo {author} {\bibfnamefont {A.}~\bibnamefont
  {Frenkel}},\ }\bibfield  {title} {\enquote {\bibinfo {title} {{A Review of
  Derivations of the Space-Time Foam Formulas}},}\ }\href@noop {} {\  (\bibinfo
  {year} {2010})},\ \Eprint {http://arxiv.org/abs/1011.1833} {arXiv:1011.1833
  [quant-ph]} \BibitemShut {NoStop}%
%%CITATION = ARXIV:1011.1833;%%
\bibitem [{\citenamefont {{Gambini}}\ \emph {et~al.}(2007)\citenamefont
  {{Gambini}}, \citenamefont {{Porto}},\ and\ \citenamefont
  {{Pullin}}}]{Gambini2007pedagogical}%
  \BibitemOpen
  \bibfield  {author} {\bibinfo {author} {\bibfnamefont {R.}~\bibnamefont
  {{Gambini}}}, \bibinfo {author} {\bibfnamefont {R.~A.}\ \bibnamefont
  {{Porto}}}, \ and\ \bibinfo {author} {\bibfnamefont {J.}~\bibnamefont
  {{Pullin}}},\ }\bibfield  {title} {\enquote {\bibinfo {title} {{Fundamental
  decoherence from quantum gravity: a pedagogical review}},}\ }\href {\doibase
  10.1007/s10714-007-0451-1} {\bibfield  {journal} {\bibinfo  {journal}
  {General Relativity and Gravitation}\ }\textbf {\bibinfo {volume} {39}},\
  \bibinfo {pages} {1143--1156} (\bibinfo {year} {2007})},\ \Eprint
  {http://arxiv.org/abs/gr-qc/0603090} {gr-qc/0603090} \BibitemShut {NoStop}%
\bibitem [{\citenamefont {Gambini}\ \emph {et~al.}(2009)\citenamefont
  {Gambini}, \citenamefont {Porto}, \citenamefont {Pullin},\ and\ \citenamefont
  {Torterolo}}]{Gambini:2008ke}%
  \BibitemOpen
  \bibfield  {author} {\bibinfo {author} {\bibfnamefont {R.}~\bibnamefont
  {Gambini}}, \bibinfo {author} {\bibfnamefont {R.~A.}\ \bibnamefont {Porto}},
  \bibinfo {author} {\bibfnamefont {J.}~\bibnamefont {Pullin}}, \ and\ \bibinfo
  {author} {\bibfnamefont {S.}~\bibnamefont {Torterolo}},\ }\bibfield  {title}
  {\enquote {\bibinfo {title} {{Conditional probabilities with Dirac
  observables and the problem of time in quantum gravity}},}\ }\href {\doibase
  10.1103/PhysRevD.79.041501} {\bibfield  {journal} {\bibinfo  {journal} {Phys.
  Rev.}\ }\textbf {\bibinfo {volume} {D79}},\ \bibinfo {pages} {041501}
  (\bibinfo {year} {2009})},\ \Eprint {http://arxiv.org/abs/0809.4235}
  {arXiv:0809.4235 [gr-qc]} \BibitemShut {NoStop}%
%%CITATION = ARXIV:0809.4235;%%
\bibitem [{\citenamefont {Page}\ and\ \citenamefont
  {Wootters}(1983)}]{Page:1983uco}%
  \BibitemOpen
  \bibfield  {author} {\bibinfo {author} {\bibfnamefont {D.~N.}\ \bibnamefont
  {Page}}\ and\ \bibinfo {author} {\bibfnamefont {W.~K.}\ \bibnamefont
  {Wootters}},\ }\bibfield  {title} {\enquote {\bibinfo {title} {{Evolution
  without evolution: dynamics described by stationary observables}},}\ }\href
  {\doibase 10.1103/PhysRevD.27.2885} {\bibfield  {journal} {\bibinfo
  {journal} {Phys. Rev.}\ }\textbf {\bibinfo {volume} {D27}},\ \bibinfo {pages}
  {2885} (\bibinfo {year} {1983})}\BibitemShut {NoStop}%
%%CITATION = PHRVA,D27,2885;%%
\bibitem [{\citenamefont {Milburn}(1991)}]{MilburnPRA1991}%
  \BibitemOpen
  \bibfield  {author} {\bibinfo {author} {\bibfnamefont {G.~J.}\ \bibnamefont
  {Milburn}},\ }\bibfield  {title} {\enquote {\bibinfo {title} {Intrinsic
  decoherence in quantum mechanics},}\ }\href {\doibase
  10.1103/PhysRevA.44.5401} {\bibfield  {journal} {\bibinfo  {journal} {Phys.
  Rev. A}\ }\textbf {\bibinfo {volume} {44}},\ \bibinfo {pages} {5401--5406}
  (\bibinfo {year} {1991})}\BibitemShut {NoStop}%
\bibitem [{\citenamefont {Egusquiza}\ \emph {et~al.}(1999)\citenamefont
  {Egusquiza}, \citenamefont {Garay},\ and\ \citenamefont
  {Raya}}]{EgusquizaPRA1999}%
  \BibitemOpen
  \bibfield  {author} {\bibinfo {author} {\bibfnamefont {I.~L.}\ \bibnamefont
  {Egusquiza}}, \bibinfo {author} {\bibfnamefont {L.~J.}\ \bibnamefont
  {Garay}}, \ and\ \bibinfo {author} {\bibfnamefont {J.~M.}\ \bibnamefont
  {Raya}},\ }\bibfield  {title} {\enquote {\bibinfo {title} {Quantum evolution
  according to real clocks},}\ }\href {\doibase 10.1103/PhysRevA.59.3236}
  {\bibfield  {journal} {\bibinfo  {journal} {Phys. Rev. A}\ }\textbf {\bibinfo
  {volume} {59}},\ \bibinfo {pages} {3236--3240} (\bibinfo {year}
  {1999})}\BibitemShut {NoStop}%
\bibitem [{\citenamefont {Di{\'o}si}(2005)}]{diosi2005intrinsic}%
  \BibitemOpen
  \bibfield  {author} {\bibinfo {author} {\bibfnamefont {L.}~\bibnamefont
  {Di{\'o}si}},\ }\bibfield  {title} {\enquote {\bibinfo {title} {Intrinsic
  time-uncertainties and decoherence: comparison of 4 models},}\ }\href@noop {}
  {\bibfield  {journal} {\bibinfo  {journal} {Brazilian Journal of Physics}\
  }\textbf {\bibinfo {volume} {35}},\ \bibinfo {pages} {260--265} (\bibinfo
  {year} {2005})}\BibitemShut {NoStop}%
\bibitem [{\citenamefont {Gambini}\ \emph {et~al.}(2006)\citenamefont
  {Gambini}, \citenamefont {Porto},\ and\ \citenamefont
  {Pullin}}]{GPPIntJModPhys2006}%
  \BibitemOpen
  \bibfield  {author} {\bibinfo {author} {\bibfnamefont {R.}~\bibnamefont
  {Gambini}}, \bibinfo {author} {\bibfnamefont {R.~A.}\ \bibnamefont {Porto}},
  \ and\ \bibinfo {author} {\bibfnamefont {J.}~\bibnamefont {Pullin}},\
  }\bibfield  {title} {\enquote {\bibinfo {title} {Fundamental spatio-temporal
  decoherence: A key to solving the conceptual problems of black holes,
  cosmology and quantum mechanics},}\ }\href@noop {} {\bibfield  {journal}
  {\bibinfo  {journal} {International Journal of Modern Physics D}\ }\textbf
  {\bibinfo {volume} {15}},\ \bibinfo {pages} {2181--2185} (\bibinfo {year}
  {2006})}\BibitemShut {NoStop}%
\bibitem [{\citenamefont {Pusey}(2018)}]{pusey2018inconsistent}%
  \BibitemOpen
  \bibfield  {author} {\bibinfo {author} {\bibfnamefont {M.~F.}\ \bibnamefont
  {Pusey}},\ }\bibfield  {title} {\enquote {\bibinfo {title} {An inconsistent
  friend},}\ }\href {https://www.nature.com/articles/s41567-018-0293-7}
  {\bibfield  {journal} {\bibinfo  {journal} {Nature Physics}\ }\textbf
  {\bibinfo {volume} {14}},\ \bibinfo {pages} {977--978} (\bibinfo {year}
  {2018})}\BibitemShut {NoStop}%
\bibitem [{\citenamefont {{Ara\'ujo}}(2016)}]{BlogMateus}%
  \BibitemOpen
  \bibfield  {author} {\bibinfo {author} {\bibfnamefont {M.}~\bibnamefont
  {{Ara\'ujo}}},\ }\bibfield  {title} {\enquote {\bibinfo {title} {If your
  interpretation of quantum mechanics has a single world but no collapse, you
  have a problem},}\ }\href
  {http://mateusaraujo.info/2016/06/20/if-your-interpretation-of-quantum-mechanics-has-a-single-world-but-no-collapse-you-have-a-problem}
  {\bibfield  {journal} {\bibinfo  {journal} {Blog \emph{More Quantum}}\ }
  (\bibinfo {year} {2016})}\BibitemShut {NoStop}%
\bibitem [{\citenamefont {{Nurgalieva}}\ and\ \citenamefont {{del
  Rio}}(2018)}]{delRio2018arXiv}%
  \BibitemOpen
  \bibfield  {author} {\bibinfo {author} {\bibfnamefont {N.}~\bibnamefont
  {{Nurgalieva}}}\ and\ \bibinfo {author} {\bibfnamefont {L.}~\bibnamefont
  {{del Rio}}},\ }\bibfield  {title} {\enquote {\bibinfo {title} {{Inadequacy
  of modal logic in quantum settings}},}\ }\href@noop {} {\bibfield  {journal}
  {\bibinfo  {journal} {ArXiv e-prints}\ } (\bibinfo {year} {2018})},\ \Eprint
  {http://arxiv.org/abs/1804.01106} {arXiv:1804.01106 [quant-ph]} \BibitemShut
  {NoStop}%
\bibitem [{\citenamefont {{Aaronson}}(2018)}]{BlogAaronson}%
  \BibitemOpen
  \bibfield  {author} {\bibinfo {author} {\bibfnamefont {S.}~\bibnamefont
  {{Aaronson}}},\ }\bibfield  {title} {\enquote {\bibinfo {title} {It’s hard
  to think when someone hadamards your brain},}\ }\href
  {https://www.scottaaronson.com/blog/?p=3975} {\bibfield  {journal} {\bibinfo
  {journal} {Blog \emph{Shtetl Optimized}}\ } (\bibinfo {year}
  {2018})}\BibitemShut {NoStop}%
\bibitem [{\citenamefont {Bub}(2018)}]{Bub2018}%
  \BibitemOpen
  \bibfield  {author} {\bibinfo {author} {\bibfnamefont {J.}~\bibnamefont
  {Bub}},\ }\bibfield  {title} {\enquote {\bibinfo {title} {{In Defense of a
  Single-World Interpretation of Quantum Mechanics}},}\ }\href@noop {}
  {\bibfield  {journal} {\bibinfo  {journal} {ArXiv e-prints}\ } (\bibinfo
  {year} {2018})},\ \Eprint {http://arxiv.org/abs/1804.03267} {arXiv:1804.03267
  [quant-ph]} \BibitemShut {NoStop}%
\bibitem [{\citenamefont {Everett~III}(1957)}]{everett1957relative}%
  \BibitemOpen
  \bibfield  {author} {\bibinfo {author} {\bibfnamefont {H.}~\bibnamefont
  {Everett~III}},\ }\bibfield  {title} {\enquote {\bibinfo {title} {" relative
  state" formulation of quantum mechanics},}\ }\href@noop {} {\bibfield
  {journal} {\bibinfo  {journal} {Reviews of modern physics}\ }\textbf
  {\bibinfo {volume} {29}},\ \bibinfo {pages} {454} (\bibinfo {year}
  {1957})}\BibitemShut {NoStop}%
\bibitem [{\citenamefont {DeWitt}(1970)}]{dewitt1970quantum}%
  \BibitemOpen
  \bibfield  {author} {\bibinfo {author} {\bibfnamefont {B.~S.}\ \bibnamefont
  {DeWitt}},\ }\bibfield  {title} {\enquote {\bibinfo {title} {Quantum
  mechanics and reality},}\ }\href@noop {} {\bibfield  {journal} {\bibinfo
  {journal} {Physics today}\ }\textbf {\bibinfo {volume} {23}},\ \bibinfo
  {pages} {30--35} (\bibinfo {year} {1970})}\BibitemShut {NoStop}%
\bibitem [{\citenamefont {Vaidman}(2016)}]{Lev-manyworlds-encyclopedia}%
  \BibitemOpen
  \bibfield  {author} {\bibinfo {author} {\bibfnamefont {L.}~\bibnamefont
  {Vaidman}},\ }\bibfield  {title} {\enquote {\bibinfo {title} {Many-worlds
  interpretation of quantum mechanics},}\ }in\ \href@noop {} {\emph {\bibinfo
  {booktitle} {The Stanford Encyclopedia of Philosophy}}},\ \bibinfo {editor}
  {edited by\ \bibinfo {editor} {\bibfnamefont {Edward~N.}\ \bibnamefont
  {Zalta}}}\ (\bibinfo  {publisher} {Metaphysics Research Lab, Stanford
  University},\ \bibinfo {year} {2016})\ \bibinfo {edition} {fall 2016}\
  ed.\BibitemShut {Stop}%
\bibitem [{\citenamefont {Fuchs}\ and\ \citenamefont
  {Schack}(2013)}]{FuchsRevModPhys2013}%
  \BibitemOpen
  \bibfield  {author} {\bibinfo {author} {\bibfnamefont {C.~A.}\ \bibnamefont
  {Fuchs}}\ and\ \bibinfo {author} {\bibfnamefont {R.}~\bibnamefont {Schack}},\
  }\bibfield  {title} {\enquote {\bibinfo {title} {Quantum-bayesian
  coherence},}\ }\href {\doibase 10.1103/RevModPhys.85.1693} {\bibfield
  {journal} {\bibinfo  {journal} {Rev. Mod. Phys.}\ }\textbf {\bibinfo {volume}
  {85}},\ \bibinfo {pages} {1693--1715} (\bibinfo {year} {2013})}\BibitemShut
  {NoStop}%
\bibitem [{\citenamefont {{Fuchs}}\ \emph {et~al.}(2014)\citenamefont
  {{Fuchs}}, \citenamefont {{Mermin}},\ and\ \citenamefont
  {{Schack}}}]{FuchsAmJPh2014}%
  \BibitemOpen
  \bibfield  {author} {\bibinfo {author} {\bibfnamefont {C.~A.}\ \bibnamefont
  {{Fuchs}}}, \bibinfo {author} {\bibfnamefont {N.~D.}\ \bibnamefont
  {{Mermin}}}, \ and\ \bibinfo {author} {\bibfnamefont {R.}~\bibnamefont
  {{Schack}}},\ }\bibfield  {title} {\enquote {\bibinfo {title} {{An
  introduction to QBism with an application to the locality of quantum
  mechanics}},}\ }\href {\doibase 10.1119/1.4874855} {\bibfield  {journal}
  {\bibinfo  {journal} {American Journal of Physics}\ }\textbf {\bibinfo
  {volume} {82}},\ \bibinfo {pages} {749--754} (\bibinfo {year} {2014})},\
  \Eprint {http://arxiv.org/abs/1311.5253} {arXiv:1311.5253 [quant-ph]}
  \BibitemShut {NoStop}%
\bibitem [{\citenamefont {Diósi}(1984)}]{Diosi1984}%
  \BibitemOpen
  \bibfield  {author} {\bibinfo {author} {\bibfnamefont {L.}~\bibnamefont
  {Diósi}},\ }\bibfield  {title} {\enquote {\bibinfo {title} {Gravitation and
  quantum-mechanical localization of macro-objects},}\ }\href {\doibase
  https://doi.org/10.1016/0375-9601(84)90397-9} {\bibfield  {journal} {\bibinfo
   {journal} {Physics Letters A}\ }\textbf {\bibinfo {volume} {105}},\ \bibinfo
  {pages} {199 -- 202} (\bibinfo {year} {1984})}\BibitemShut {NoStop}%
\bibitem [{\citenamefont {Ghirardi}\ \emph {et~al.}(1986)\citenamefont
  {Ghirardi}, \citenamefont {Rimini},\ and\ \citenamefont {Weber}}]{GRW1986}%
  \BibitemOpen
  \bibfield  {author} {\bibinfo {author} {\bibfnamefont {G.~C.}\ \bibnamefont
  {Ghirardi}}, \bibinfo {author} {\bibfnamefont {A.}~\bibnamefont {Rimini}}, \
  and\ \bibinfo {author} {\bibfnamefont {T.}~\bibnamefont {Weber}},\ }\bibfield
   {title} {\enquote {\bibinfo {title} {Unified dynamics for microscopic and
  macroscopic systems},}\ }\href@noop {} {\bibfield  {journal} {\bibinfo
  {journal} {Physical Review D}\ }\textbf {\bibinfo {volume} {34}},\ \bibinfo
  {pages} {470} (\bibinfo {year} {1986})}\BibitemShut {NoStop}%
\bibitem [{\citenamefont {Di{\'o}si}(1988)}]{diosi1988continuous}%
  \BibitemOpen
  \bibfield  {author} {\bibinfo {author} {\bibfnamefont {L.}~\bibnamefont
  {Di{\'o}si}},\ }\bibfield  {title} {\enquote {\bibinfo {title} {Continuous
  quantum measurement and it{\^o} formalism},}\ }\href@noop {} {\bibfield
  {journal} {\bibinfo  {journal} {Physics Letters A}\ }\textbf {\bibinfo
  {volume} {129}},\ \bibinfo {pages} {419--423} (\bibinfo {year}
  {1988})}\BibitemShut {NoStop}%
\bibitem [{\citenamefont {Pearle}(1989)}]{PearlePRA1989}%
  \BibitemOpen
  \bibfield  {author} {\bibinfo {author} {\bibfnamefont {P.}~\bibnamefont
  {Pearle}},\ }\bibfield  {title} {\enquote {\bibinfo {title} {Combining
  stochastic dynamical state-vector reduction with spontaneous localization},}\
  }\href {\doibase 10.1103/PhysRevA.39.2277} {\bibfield  {journal} {\bibinfo
  {journal} {Phys. Rev. A}\ }\textbf {\bibinfo {volume} {39}},\ \bibinfo
  {pages} {2277--2289} (\bibinfo {year} {1989})}\BibitemShut {NoStop}%
\bibitem [{\citenamefont {Ghirardi}\ \emph {et~al.}(1990)\citenamefont
  {Ghirardi}, \citenamefont {Pearle},\ and\ \citenamefont
  {Rimini}}]{GhirardiPRA1990}%
  \BibitemOpen
  \bibfield  {author} {\bibinfo {author} {\bibfnamefont {G.~C.}\ \bibnamefont
  {Ghirardi}}, \bibinfo {author} {\bibfnamefont {P.}~\bibnamefont {Pearle}}, \
  and\ \bibinfo {author} {\bibfnamefont {A.}~\bibnamefont {Rimini}},\
  }\bibfield  {title} {\enquote {\bibinfo {title} {Markov processes in hilbert
  space and continuous spontaneous localization of systems of identical
  particles},}\ }\href {\doibase 10.1103/PhysRevA.42.78} {\bibfield  {journal}
  {\bibinfo  {journal} {Phys. Rev. A}\ }\textbf {\bibinfo {volume} {42}},\
  \bibinfo {pages} {78--89} (\bibinfo {year} {1990})}\BibitemShut {NoStop}%
\bibitem [{\citenamefont {Gisin}(1989)}]{gisin1989stochastic}%
  \BibitemOpen
  \bibfield  {author} {\bibinfo {author} {\bibfnamefont {N.}~\bibnamefont
  {Gisin}},\ }\bibfield  {title} {\enquote {\bibinfo {title} {Stochastic
  quantum dynamics and relativity},}\ }\href@noop {} {\bibfield  {journal}
  {\bibinfo  {journal} {Helv. Phys. Acta}\ }\textbf {\bibinfo {volume} {62}},\
  \bibinfo {pages} {363--371} (\bibinfo {year} {1989})}\BibitemShut {NoStop}%
\bibitem [{\citenamefont {Penrose}(1996)}]{penrose1996gravity}%
  \BibitemOpen
  \bibfield  {author} {\bibinfo {author} {\bibfnamefont {R.}~\bibnamefont
  {Penrose}},\ }\bibfield  {title} {\enquote {\bibinfo {title} {On gravity's
  role in quantum state reduction},}\ }\href@noop {} {\bibfield  {journal}
  {\bibinfo  {journal} {General relativity and gravitation}\ }\textbf {\bibinfo
  {volume} {28}},\ \bibinfo {pages} {581--600} (\bibinfo {year}
  {1996})}\BibitemShut {NoStop}%
\bibitem [{\citenamefont {{Gisin}}(2017)}]{Gisinarxiv2017}%
  \BibitemOpen
  \bibfield  {author} {\bibinfo {author} {\bibfnamefont {N.}~\bibnamefont
  {{Gisin}}},\ }\bibfield  {title} {\enquote {\bibinfo {title} {{Collapse. What
  else?}}}\ }\href@noop {} {\bibfield  {journal} {\bibinfo  {journal} {ArXiv
  e-prints}\ } (\bibinfo {year} {2017})},\ \Eprint
  {http://arxiv.org/abs/1701.08300} {arXiv:1701.08300 [quant-ph]} \BibitemShut
  {NoStop}%
\bibitem [{\citenamefont {{Mueller}}(2017)}]{MuellerarXiv2017}%
  \BibitemOpen
  \bibfield  {author} {\bibinfo {author} {\bibfnamefont {M.~P.}\ \bibnamefont
  {{Mueller}}},\ }\bibfield  {title} {\enquote {\bibinfo {title} {{Could the
  physical world be emergent instead of fundamental, and why should we ask?
  (short version)}},}\ }\href@noop {} {\bibfield  {journal} {\bibinfo
  {journal} {ArXiv e-prints}\ } (\bibinfo {year} {2017})},\ \Eprint
  {http://arxiv.org/abs/1712.01816} {arXiv:1712.01816 [quant-ph]} \BibitemShut
  {NoStop}%
\bibitem [{\citenamefont {Fr\"owis}\ \emph {et~al.}(2018)\citenamefont
  {Fr\"owis}, \citenamefont {Sekatski}, \citenamefont {D\"ur}, \citenamefont
  {Gisin},\ and\ \citenamefont {Sangouard}}]{FrowisRevModPhys2018}%
  \BibitemOpen
  \bibfield  {author} {\bibinfo {author} {\bibfnamefont {F.}~\bibnamefont
  {Fr\"owis}}, \bibinfo {author} {\bibfnamefont {P.}~\bibnamefont {Sekatski}},
  \bibinfo {author} {\bibfnamefont {W.}~\bibnamefont {D\"ur}}, \bibinfo
  {author} {\bibfnamefont {N.}~\bibnamefont {Gisin}}, \ and\ \bibinfo {author}
  {\bibfnamefont {N.}~\bibnamefont {Sangouard}},\ }\bibfield  {title} {\enquote
  {\bibinfo {title} {Macroscopic quantum states: Measures, fragility, and
  implementations},}\ }\href {\doibase 10.1103/RevModPhys.90.025004} {\bibfield
   {journal} {\bibinfo  {journal} {Rev. Mod. Phys.}\ }\textbf {\bibinfo
  {volume} {90}},\ \bibinfo {pages} {025004} (\bibinfo {year}
  {2018})}\BibitemShut {NoStop}%
\bibitem [{\citenamefont {Vinante}\ \emph {et~al.}(2017)\citenamefont
  {Vinante}, \citenamefont {Mezzena}, \citenamefont {Falferi}, \citenamefont
  {Carlesso},\ and\ \citenamefont {Bassi}}]{BassiPRL2017}%
  \BibitemOpen
  \bibfield  {author} {\bibinfo {author} {\bibfnamefont {A.}~\bibnamefont
  {Vinante}}, \bibinfo {author} {\bibfnamefont {R.}~\bibnamefont {Mezzena}},
  \bibinfo {author} {\bibfnamefont {P.}~\bibnamefont {Falferi}}, \bibinfo
  {author} {\bibfnamefont {M.}~\bibnamefont {Carlesso}}, \ and\ \bibinfo
  {author} {\bibfnamefont {A.}~\bibnamefont {Bassi}},\ }\bibfield  {title}
  {\enquote {\bibinfo {title} {Improved noninterferometric test of collapse
  models using ultracold cantilevers},}\ }\href {\doibase
  10.1103/PhysRevLett.119.110401} {\bibfield  {journal} {\bibinfo  {journal}
  {Phys. Rev. Lett.}\ }\textbf {\bibinfo {volume} {119}},\ \bibinfo {pages}
  {110401} (\bibinfo {year} {2017})}\BibitemShut {NoStop}%
\end{thebibliography}%
\end{document}